\documentclass[12pt,letterpaper]{article}%
\usepackage{amsmath}
\usepackage{amsthm}
\usepackage{amssymb}
\usepackage{amsfonts}
\usepackage{mathpazo}
\usepackage{times}
\usepackage[T1]{fontenc}
\usepackage[authoryear]{natbib}
\usepackage{titlesec}
\usepackage{xcolor}
\usepackage{hyperref}
\usepackage[normalem]{ulem}
\usepackage{graphicx}
\usepackage[tight]{subfigure}
\usepackage{epstopdf}
\usepackage{caption}%
\providecommand{\U}[1]{\protect\rule{.1in}{.1in}}
\makeatletter
\long
\def\@makecaption#1#2{  \vskip\abovecaptionskip
\sbox\@tempboxa{{\captionfonts #1: #2}}  \ifdim \wd\@tempboxa >\hsize
{\captionfonts #1: #2\par}
\else
\hbox to\hsize{\hfil\box\@tempboxa\hfil}  \fi
\vskip\belowcaptionskip}
\makeatother
\makeatletter
\@addtoreset{equation}{section}
\makeatother
\renewcommand{\theequation}{{\thesection}.{\arabic{equation}}}
\titleformat{\subsection}{\normalfont\bfseries}{\thesubsection}{0.5em}{}
\titleformat{\subsubsection}{\normalfont\bfseries\itshape}{\thesubsubsection}{0.5em}{}
\titleformat{\paragraph}[runin]{\normalfont\bfseries}{\theparagraph}{0.5em}{}[.]
\renewcommand{\thesubsubsection}{\arabic{section}.\arabic{subsection}.\arabic{subsubsection}}
\renewcommand{\theparagraph}{(\arabic{paragraph})}
\newtheoremstyle{thm}{1.0em}{1.0em}{\itshape}{}{\bfseries}{.}{.5em}{}
\theoremstyle{thm}
\newtheorem{theorem}{Theorem}[section]

\newtheoremstyle{rem}{1.0em}{1.0em}{\normalfont}{}{\bfseries}{.}{.5em}{}
\theoremstyle{rem}
\newtheorem{remark}{Remark}[section]

\newenvironment{proof}[1][Proof]{\par\vspace{1.0em}\noindent\textbf{#1. }}{\hfill $\Box$\par\vspace{1.0em}}
\hypersetup{
bookmarks = true,
bookmarksnumbered = true,
bookmarksopen = true,
bookmarksopenlevel = 3,
pdfstartview = FitH,
colorlinks = false,
citebordercolor = {white},
linkbordercolor = {red}
}
\newcommand{\captionfonts}{\small}

\providecommand{\argmin}{\mathop{\mathrm{arg\,min\,}}}

\renewcommand{\qed}{\hfill $\Box$}
\AtBeginDocument{
\DeclareSymbolFont{AMSb}{U}{msb}{m}{n}
\DeclareSymbolFontAlphabet{\mathbb}{AMSb}
}
\ifx\pdfoutput\relax\let\pdfoutput=\undefined\fi
\newcount\msipdfoutput
\ifx\pdfoutput\undefined\else
\ifcase\pdfoutput\else
\ifx\paperwidth\undefined\else
\ifdim\paperheight=0pt\relax\else\pdfpageheight\paperheight\fi
\ifdim\paperwidth=0pt\relax\else\pdfpagewidth\paperwidth\fi
\fi\fi\fi
\begin{document}
\hspace{13.9cm}1

\ \vspace{1mm}\newline

\noindent{\Large \textbf{Approximations of Shannon Mutual Information for Discrete Variables with Applications to Neural Population Coding}}

\ \newline\textbf{{\large Wentao Huang}}\newline\textit{wnthuang@gmail.com}
\newline\textit{Key Laboratory of Cognition and Intelligence and Information
Science Academy of China Electronics Technology Group Corporation, Beijing
100086, China, Department of Biomedical Engineering, Johns Hopkins University
School of Medicine, Baltimore, MD 21205, U.S.A.}
\medskip\newline\textbf{{\large Kechen Zhang}}\newline%
\textit{kzhang4@jhmi.edu\newline Department of Biomedical Engineering, Johns
Hopkins University School of Medicine, Baltimore, MD 21205, U.S.A.}\newline

\begin{center}
\textbf{Abstract}
\end{center}
\textbf{Although Shannon mutual information has been widely used, its effective
calculation is often difficult for many practical problems, including those in
neural population coding. Asymptotic formulas based on Fisher information
sometimes provide accurate approximations to the mutual information but this
approach is restricted to continuous variables because the calculation of
Fisher information requires derivatives with respect to the encoded variables.
In this paper, we consider information-theoretic bounds and approximations of
the mutual information based on Kullback--Leibler divergence and R\'{e}nyi
divergence. We propose several information metrics to approximate Shannon
mutual information in the context of neural population coding. While our
asymptotic formulas all work for discrete variables, one of them has
consistent performance and high accuracy regardless of whether the encoded
variables are discrete or continuous. We performed numerical simulations and
confirmed that our approximation formulas were highly accurate for
approximating the mutual information between the stimuli and the responses of
a large neural population. These approximation formulas may potentially bring
convenience to the applications of information theory to many practical and
theoretical problems.}

\section{Introduction}

\label{Sec:1} Information theory is a powerful tool widely used in many
disciplines, including, for example, neuroscience, machine learning, and
communication technology
\citep{Borst(1999-information), Pouget(2000-information),
Laughlin(2003-communication), Brown(2004-multiple), Bell(1997-independent),
Huang2017-IC-informationa, Huang2017-IC-information}. As it is often
notoriously difficult to effectively calculate Shannon mutual information in
many practical applications~\citep{Cover(2006-BK-elements)}, various
approximation methods have been proposed to estimate the mutual information,
such as those based on asymptotic expansion \citep{Miller(1955-note),
Carlton(1969-bias), Treves(1995-upward), Victor(2000-asymptotic),
Paninski(2003-estimation)}, \textit{k}-nearest
neighbor~\citep{Kraskov(2004-estimating)}, and minimal spanning trees
\citep{Khan(2007-relative)}. Recently, Safaai~{et~al.}\ proposed a copula
method for estimation of mutual information, which can be nonparametric and
potentially robust~\citep{Safaai2018-information}. Another approach for
estimating the mutual information is to simplify the calculations by
approximations based on information-theoretic bounds, such as the
Cram\'{e}r--Rao lower bound~\citep{Rao(1945-information)} and the van Trees'
Bayesian Cram\'{e}r--Rao bound~\citep{VanTrees(2007-BK-bayesian)}.

{In this paper, we focus on mutual information estimation based on asymptotic
approximations~\citep{Clarke(1990-information), Rissanen(1996-fisher), Brunel(1998-mutual), Sompolinsky(2001-population), Kang(2001-mutual), Huang(2017-information)}%
.} For encoding of continuous variables, asymptotic relations between mutual
information and Fisher information have been presented by several researchers
\citep{Clarke(1990-information), Rissanen(1996-fisher), Brunel(1998-mutual), Sompolinsky(2001-population)}%
. Recently, Huang and Zhang~\citep{Huang(2017-information)} proposed an
improved approximation formula, which remains accurate for high-dimensional
variables. A significant advantage of this approach is that asymptotic
approximations are sometimes very useful in analytical studies. For instance,
asymptotic approximations allow us to prove that the optimal neural population
distribution that maximizes the mutual information between stimulus and
response can be solved by convex optimization~\citep{Huang(2017-information)}.
Unfortunately this approach does not generalize to discrete variables since
the calculation of Fisher information requires partial derivatives of the
likelihood function with respect to the encoded variables. For encoding of
discrete variables, Kang and Sompolinsky \citep{Kang(2001-mutual)} represented
an asymptotic relationship between mutual information and Chernoff information
for statistically independent neurons in a large population. However, Chernoff
information is still hard to calculate in many practical applications.

Discrete stimuli or variables occur naturally in sensory coding. While some
stimuli are continuous (e.g., the direction of movement, and the pitch of a
tone), others are discrete (e.g., the identities of faces, and the words in
human speech). For definiteness, in this paper, we frame our questions in the
context of neural population coding; that is, we assume that the stimuli or
the input variables are encoded by the pattern of responses elicited from a
large population of neurons. The concrete examples used in our numerical
simulations were based on Poisson spike model, where the response of each
neuron is taken as the spike count within a given time window. While this
simple Poisson model allowed us to consider a large neural population, it only
captured the spike rate but not any temporal structure of the spike
trains~\citep{Strong1998-entropy, Nemenman2004-entropy, Panzeri2017-correcting,
Houghton2019-calculating}. Nonetheless, our mathematical results are quite
general and should be applicable to other input--output systems under suitable
conditions to be discussed later.

In the following, we first derive several upper and lower bounds on Shannon
mutual information using Kullback--Leibler divergence and R\'{e}nyi
divergence. Next, we derive several new approximation formulas for Shannon
mutual information in the limit of large population size. These formulas are
more convenient to calculate than the mutual information in our examples.
Finally, we confirm the validity of our approximation formulas using the true
mutual information as evaluated by Monte Carlo simulations.

\section{Theory and Methods}

\label{Sec:2} \vspace{-6pt}

\subsection{ Notations and Definitions}

\label{Sec:2.1}

Suppose the input $\mathbf{x}$ is a $K$-dimensional vector, $\mathbf{x}%
=(x_{1},\cdots,x_{K})^{T}$, which could be interpreted as the parameters that
specifies a stimulus for a sensory system, and the outputs is an
$N$-dimensional vector, $\mathbf{r}=(r_{1},\cdots,r_{N})^{T}$, which could be
interpreted as the responses of $N$ neurons. We assume $N$ is large, generally
$N\gg K$. We denote random variables by upper case letters, e.g., random
variables $X$ and $R$, in contrast to their vector values $\mathbf{x}$ and
$\mathbf{r}$. The mutual information between $X$ and $R$ is defined~by
\begin{equation}
I=I(X;R)=\left\langle \ln\frac{p(\mathbf{r}|\mathbf{x})}{p(\mathbf{r}%
)}\right\rangle _{\mathbf{r},\mathbf{x}}, \label{MI}%
\end{equation}
where $\mathbf{x}\in{{\mathcal{X}}}\subseteq%
%TCIMACRO{\U{211d} }%
%BeginExpansion
\mathbb{R}
%EndExpansion
^{K}$, $\mathbf{r}\in\mathcal{R}\subseteq%
%TCIMACRO{\U{211d} }%
%BeginExpansion
\mathbb{R}
%EndExpansion
^{N}$, and $\left\langle \cdot\right\rangle _{\mathbf{r},\mathbf{x}}$ denotes
the expectation with respect to the probability density function
$p(\mathbf{r},\mathbf{x})$. Similarly, in the following, we use $\left\langle
\cdot\right\rangle _{\mathbf{r}|\mathbf{x}}$ and $\left\langle \cdot
\right\rangle _{\mathbf{x}}$ to denote expectations with respect to
$p(\mathbf{r}|\mathbf{x})$ and $p(\mathbf{x})$, respectively.

If $p(\mathbf{x})$ and $p(\mathbf{r}|\mathbf{x})$ are twice continuously
differentiable for almost every $\mathbf{x}\in{{\mathcal{X}}}$, then for large
$N$ we can use an asymptotic formula to approximate the true value of $I$ with
high accuracy~\citep{Huang(2017-information)}:
\begin{equation}
I\simeq I_{G}=\frac{1}{2}\left\langle \ln\left(  \det\left(  \frac
{\mathbf{G}(\mathbf{x})}{2\pi e}\right)  \right)  \right\rangle _{\mathbf{x}%
}+H(X), \label{Ia}%
\end{equation}
which is sometimes reduced to%
\begin{equation}
I\simeq I_{F}=\frac{1}{2}\left\langle \ln\left(  \det\left(  \frac
{\mathbf{J}(\mathbf{x})}{2\pi e}\right)  \right)  \right\rangle _{\mathbf{x}%
}+H(X), \label{Ib}%
\end{equation}
where $\det\left(  \cdot\right)  $ denotes the matrix determinant,
$H(X)=-\left\langle \ln p(\mathbf{x})\right\rangle _{\mathbf{x}}$ is the
stimulus entropy,
\begin{align}
&  \mathbf{G}(\mathbf{x})=\mathbf{J}(\mathbf{x})+\mathbf{P}\left(
\mathbf{x}\right)  ,\label{Gx}\\
&  \mathbf{P}(\mathbf{x})=-{\dfrac{\partial^{2}\ln p(\mathbf{x})}%
{\partial\mathbf{x}\partial\mathbf{x}^{T}}}, \label{Px}%
\end{align}
and
\begin{equation}
\mathbf{J}(\mathbf{x})=-\left\langle \frac{\partial^{2}\ln p(\mathbf{r}%
|\mathbf{x})}{\partial\mathbf{x}\partial\mathbf{x}^{T}}\right\rangle
_{\mathbf{r}|\mathbf{x}} = \left\langle \frac{\partial\ln p(\mathbf{r}%
|\mathbf{x})}{\partial\mathbf{x}}\frac{\partial\ln p(\mathbf{r}|\mathbf{x}%
)}{\partial\mathbf{x}^{T}}\right\rangle _{\mathbf{r}|\mathbf{x}} \label{Jx}%
\end{equation}
is the Fisher information matrix.

We denote the Kullback--Leibler divergence as
\begin{equation}
D\left(  \mathbf{x}||{\hat{\mathbf{x}}}\right)  =\left\langle \ln
\frac{p\left(  \mathbf{r}|\mathbf{x}\right)  }{p\left(  \mathbf{r}%
|{\hat{\mathbf{x}}}\right)  }\right\rangle _{\mathbf{r}|\mathbf{x}},
\label{DKL}%
\end{equation}
and denote R{\'{e}}nyi divergence~\citep{Renyi(1961-IP-measures)} of order
$\beta+1$ as
\begin{equation}
D_{\beta}\left(  \mathbf{x}||{\hat{\mathbf{x}}}\right)  =\frac{1}{\beta}%
\ln\left\langle \left(  \frac{p(\mathbf{r}|{{\mathbf{x}}})}{p\left(
\mathbf{r}|\hat{\mathbf{x}}\right)  }\right)  ^{\beta}\right\rangle
_{\mathbf{r}|\mathbf{x}}. \label{DRenyi}%
\end{equation}

Here, $\beta D_{\beta}\left(  \mathbf{x}||{\hat{\mathbf{x}}}\right)  $ is
equivalent to Chernoff divergence of order $\beta+1$
\citep{chernoff1952measure}. It is well known that $D_{\beta}\left(
\mathbf{x}||{\hat{\mathbf{x}}}\right)  \to D\left(  \mathbf{x}||{\hat
{\mathbf{x}}}\right)  $ in the limit $\beta\to0$.

We define%
\begin{align}
I_{u}  &  =-\left\langle \ln\left\langle \exp\left(  -D\left(  \mathbf{x}%
||{\hat{\mathbf{x}}}\right)  \right)  \right\rangle _{{\hat{\mathbf{x}}}%
}\right\rangle _{\mathbf{x}},\label{Iu}\\
I_{e}  &  =-\left\langle \ln\left\langle \exp\left(  -e^{-1}D\left(
\mathbf{x}||{\hat{\mathbf{x}}}\right)  \right)  \right\rangle _{{\hat
{\mathbf{x}}}}\right\rangle _{\mathbf{x}},\label{Ie}\\
I_{\beta,\alpha}  &  =-\left\langle \ln\left\langle \exp\left(  -\beta
D_{\beta}\left(  \mathbf{x}||{\hat{\mathbf{x}}}\right)  +\left(
1-\alpha\right)  \ln\frac{p\left(  \mathbf{x}\right)  }{p\left(
{\hat{\mathbf{x}}}\right)  }\right)  \right\rangle _{{\hat{\mathbf{x}}}%
}\right\rangle _{\mathbf{x}}, \label{Ibeta}%
\end{align}
where in $I_{\beta,\alpha}$ we have $\beta\in\left(  0\text{,\thinspace
}1\right)  $ and $\alpha\in\left(  0\text{,\thinspace}\infty\right)  $ and
assume $p\left(  \mathbf{x}\right)  >0$ for all $\mathbf{x}\in{{\mathcal{X}}}$.

In the following, we suppose $\mathbf{x}$ takes $M$ discrete values,
$\mathbf{x}_{m}$, $m\in\mathcal{M}=\left\{  1\text{,\thinspace}%
2\text{,\thinspace}\cdots\text{,\thinspace}M\right\}  $, and $p(\mathbf{x}%
_{m})>0$ for all $m$. Now, the definitions in Equations (\ref{Iu}%
)--(\ref{Ibeta}) become
\begin{align}
I_{u}  &  =-\sum\limits_{m=1}^{M}p\left(  \mathbf{x}_{m}\right)  \ln\left(
\sum\limits_{\hat{m}=1}^{M}\frac{p\left(  \mathbf{x}_{\hat{m}}\right)
}{p\left(  \mathbf{x}_{m}\right)  }\exp\left(  -D\left(  \mathbf{x}%
_{m}\mathbf{||x}_{\hat{m}}\right)  \right)  \right)  +H(X),\label{Rem1.Iu}\\
I_{e}  &  =-\sum\limits_{m=1}^{M}p\left(  \mathbf{x}_{m}\right)  \ln\left(
\sum\limits_{\hat{m}=1}^{M}\frac{p\left(  \mathbf{x}_{\hat{m}}\right)
}{p\left(  \mathbf{x}_{m}\right)  }\exp\left(  -e^{-1}D\left(  \mathbf{x}%
_{m}\mathbf{||x}_{\hat{m}}\right)  \right)  \right)  +H(X),\label{Rem1.Ie}\\
I_{\beta,\alpha}  &  =-\sum\limits_{m=1}^{M}p\left(  \mathbf{x}_{m}\right)
\ln\left(  \sum\limits_{\hat{m}=1}^{M}\left(  \frac{p(\mathbf{x}_{{\hat{m}}}%
)}{p(\mathbf{x}_{m})}\right)  ^{\alpha}\exp\left(  -\beta D_{\beta}\left(
\mathbf{x}_{m}||\mathbf{x}_{\hat{m}}\right)  \right)  \right)  +H(X).
\label{Rem1.Ibeta}%
\end{align}

Furthermore, we define
\begin{align}
I_{d}  &  =-\sum\limits_{m=1}^{M}p(\mathbf{x}_{m})\ln\left(  1+\sum
\limits_{{\hat{m}}\in{\mathcal{M}}_{m}^{u}}\frac{p(\mathbf{x}_{{\hat{m}}}%
)}{p(\mathbf{x}_{m})}\exp\left(  -e^{-1}D\left(  \mathbf{x}_{m}||\mathbf{x}%
_{{\hat{m}}}\right)  \right)  \right)  +H\left(  X\right)  ,\label{Id}\\
I_{u}^{d}  &  =-\sum\limits_{m=1}^{M}p(\mathbf{x}_{m})\ln\left(
1+\sum\limits_{{\hat{m}}\in{\mathcal{M}}_{m}^{u}}\frac{p(\mathbf{x}_{{\hat{m}%
}})}{p(\mathbf{x}_{m})}\exp\left(  -D\left(  \mathbf{x}_{m}||\mathbf{x}%
_{{\hat{m}}}\right)  \right)  \right)  +H\left(  X\right)  ,\label{Iud}\\
I_{\beta,\alpha}^{d}  &  =-\sum\limits_{m=1}^{M}p(\mathbf{x}_{m})\ln\left(
1+\sum\limits_{{\hat{m}}\in{\mathcal{M}}_{m}^{\beta}}\left(  \frac
{p(\mathbf{x}_{{\hat{m}}})}{p(\mathbf{x}_{m})}\right)  ^{\alpha}\exp\left(
-\beta D_{\beta}\left(  \mathbf{x}_{m}||\mathbf{x}_{{\hat{m}}}\right)
\right)  \right)  +H\left(  X\right)  ,\label{Ibd}\\
I_{D}  &  =-\sum\limits_{m=1}^{M}p(\mathbf{x}_{m})\ln\left(  1+\sum
\limits_{{\hat{m}}\in{\mathcal{M}}_{m}^{u}}\exp\left(  -e^{-1}D\left(
\mathbf{x}_{m}||\mathbf{x}_{{\hat{m}}}\right)  \right)  \right)  +H\left(
X\right)  , \label{ID}%
\end{align}
where
\begin{align}
&  {\mathcal{\check{M}}}_{m}^{\beta}=\left\{  {\hat{m}}:{\hat{m}%
}={\underset{\check{m}\in{\mathcal{M}}-{\mathcal{\hat{M}}}_{m}^{\beta
}}{\argmin}}D_{\beta}\left(  \mathbf{x}_{m}||\mathbf{x}_{\check{m}}\right)
\right\}  ,\label{Thm1a.3a}\\
&  {\mathcal{\check{M}}}_{m}^{u}=\left\{  {\hat{m}}:{\hat{m}}%
={\underset{\check{m}\in{\mathcal{M}}-{\mathcal{\hat{M}}}_{m}^{u}}{\argmin}%
}D\left(  \mathbf{x}_{m}||\mathbf{x}_{\check{m}}\right)  \right\}
,\label{Thm1a.3b}\\
&  {\mathcal{\hat{M}}}_{m}^{\beta}=\left\{  {\hat{m}}:D_{\beta}\left(
\mathbf{x}_{m}||\mathbf{x}_{{\hat{m}}}\right)  =0\right\}  ,\label{Thm1a.3c}\\
&  {\mathcal{\hat{M}}}_{m}^{u}=\left\{  {\hat{m}}:D\left(  \mathbf{x}%
_{m}||\mathbf{x}_{{\hat{m}}}\right)  =0\right\}  ,\label{Thm1a.3d}\\
&  {\mathcal{M}}_{m}^{\beta}={\mathcal{\check{M}}}_{m}^{\beta}\cup
{\mathcal{\hat{M}}}_{m}^{\beta}-\left\{  m\right\}  ,\label{Thm1a.3e}\\
&  {\mathcal{M}}_{m}^{u}={\mathcal{\check{M}}}_{m}^{u}\cup{\mathcal{\hat{M}}%
}_{m}^{u}-\left\{  m\right\}  . \label{Thm1a.3f}%
\end{align}

Here, notice that, if $\mathbf{x}$ is uniformly distributed, then by
definition $I_{d}$ and $I_{D}$ become identical. The elements in set
${\mathcal{\check{M}}}_{m}^{\beta}$ are those that make $D_{\beta}\left(
\mathbf{x}_{m}||\mathbf{x}_{\check{m}}\right)  $ take the minimum value,
excluding any element that satisfies the condition $D_{\beta}\left(
\mathbf{x}_{m}||\mathbf{x}_{{\hat{m}}}\right)  =0$. Similarly, the elements in
set ${\mathcal{\check{M}}}_{m}^{u}$ are those that minimize $D\left(
\mathbf{x}_{m}||\mathbf{x}_{\check{m}}\right)  $ excluding the ones that
satisfy the condition $D\left(  \mathbf{x}_{m}||\mathbf{x}_{{\hat{m}}}\right)
=0$.

\subsection{Theorems}

In the following, we state several conclusions as theorems and prove them in Appendix.

\begin{theorem}
\label{Theorem 1} The mutual information $I$ is bounded as follows:
\begin{equation}
I_{\beta,\alpha}\leq I\leq I_{u}. \label{Lma1a}%
\end{equation}

\end{theorem}

\begin{theorem}
\label{Theorem 2} The following inequalities are satisfied,
\begin{equation}
I_{\beta_{1},1}\leq I_{e}\leq I_{u} \label{Ib1<Ie<Iu}%
\end{equation}
where $I_{\beta_{1},1}$ is a special case of $I_{\beta,\alpha}$ in
Equation~\eqref{Ibeta} with $\beta_{1}=e^{-1}$ so that
\begin{equation}
I_{\beta_{1},1}=-\left\langle \ln\left\langle \exp\left(  -\beta_{1}%
D_{\beta_{1}}\left(  \mathbf{x}||{\hat{\mathbf{x}}}\right)  \right)
\right\rangle _{{\hat{\mathbf{x}}}}\right\rangle _{\mathbf{x}}. \label{Ib1}%
\end{equation}

\end{theorem}

\begin{theorem}
\label{Theorem 3} If there exist $\gamma_{1}>0$ and $\gamma_{2}>0$ such that%
\begin{align}
\beta D_{\beta}\left(  \mathbf{x}_{m}||\mathbf{x}_{m_{1}}\right)   &
\geq\gamma_{1}\ln N,\label{Thm1a.1a}\\
D\left(  \mathbf{x}_{m}||\mathbf{x}_{m_{2}}\right)   &  \geq\gamma_{2}\ln N,
\label{Thm1a.1b}%
\end{align}
for discrete stimuli $\mathbf{x}_{m}$, where $m\in\mathcal{M}$, $m_{1}%
\in\mathcal{M}-{\mathcal{M}}_{m}^{\beta}$ and $m_{2}\in\mathcal{M}%
-{\mathcal{M}}_{m}^{u}$, then we have the following asymptotic relationships:
\begin{equation}
I_{\beta,\alpha}=I_{\beta,\alpha}^{d}+O\left(  N^{-\gamma_{1}}\right)  \leq
I\leq I_{u}=I_{u}^{d}+O\left(  N^{-\gamma_{2}}\right)  \label{Thm1a}%
\end{equation}
and%
\begin{equation}
I_{e}=I_{d}+O\left(  N^{-{\gamma_{2}}/{e}}\right)  . \label{Thm1a.0}%
\end{equation}

\end{theorem}

\begin{theorem}
\label{Theorem 4} Suppose $p(\mathbf{x})$ and $p(\mathbf{r}|\mathbf{x})$ are
twice continuously differentiable for $\mathbf{x}\in{{\mathcal{X}}}$,
${\left\Vert {q}{^{\prime}(\mathbf{x})}\right\Vert }<\infty$, ${\left\Vert
{{q}^{\prime\prime}(\mathbf{x})}\right\Vert }<\infty$, where $q(\mathbf{x}%
)=\ln p(\mathbf{x})$ and $^{\prime}$ and $^{\prime\prime}$ denote partial
derivatives $\partial/\partial\mathbf{x}$ and $\partial^{2}/\partial
\mathbf{x}\partial\mathbf{x}^{T}$, and $\mathbf{G}_{\gamma}(\mathbf{x})$ is
positive definite with ${\left\Vert N\mathbf{G}_{\gamma}^{-1}\left(
\mathbf{x}\right)  \right\Vert }=O\left(  1\right)  $, where ${\left\Vert
{\mathbf{\cdot}}\right\Vert }$ denotes matrix Frobenius norm,%
\begin{equation}
\mathbf{G}_{\gamma}(\mathbf{x})=\gamma\left(  \mathbf{J}(\mathbf{x}%
)+\mathbf{P}\left(  \mathbf{x}\right)  \right)  , \label{Thm1A.0}%
\end{equation}
$\gamma=\beta\left(  1-\beta\right)  $ and $\beta\in\left(  0\text{,\thinspace
}1\right)  $. If there exist an $\omega=\omega\left(  \mathbf{x}\right)  >0$
such that
\begin{align}
&  \det\left(  \mathbf{G}(\mathbf{x})\right)  ^{1/2}\int_{{{\mathcal{\bar{X}}%
}}_{\varepsilon}\left(  {\mathbf{x}}\right)  }p({\hat{\mathbf{x}}})\exp\left(
-D\left(  \mathbf{x}||{\hat{\mathbf{x}}}\right)  \right)  d{\hat{\mathbf{x}}%
}=O\left(  N^{-1}\right)  ,\label{Thm1A.1a}\\
&  \det\left(  \mathbf{G}_{\gamma}(\mathbf{x})\right)  ^{1/2}\int%
_{{{\mathcal{\bar{X}}}}_{\varepsilon}\left(  {\mathbf{x}}\right)  }%
p({\hat{\mathbf{x}}})\exp\left(  -\beta D_{\beta}\left(  \mathbf{x}%
||{\hat{\mathbf{x}}}\right)  \right)  d{\hat{\mathbf{x}}}=O\left(
N^{-1}\right)  , \label{Thm1A.1b}%
\end{align}
for all $\mathbf{x}\in{{\mathcal{X}}}$ and $\varepsilon\in\left(
0\text{,\thinspace}\omega\right)  $, where $\mathcal{\bar{X}}_{\omega
}(\mathbf{x})=\mathcal{X}-{{\mathcal{X}}}_{\omega}(\mathbf{x})$ is the
complementary set of ${{\mathcal{X}}}_{\omega}(\mathbf{x})=\left\{
\breve{\mathbf{x}}\in%
%TCIMACRO{\U{211d} }%
%BeginExpansion
\mathbb{R}
%EndExpansion
^{K}:\left(  \breve{\mathbf{x}}-\mathbf{x}\right)  ^{T}\mathbf{G}%
(\mathbf{x})\left(  \breve{\mathbf{x}}-\mathbf{x}\right)  <N\omega
^{2}\right\}  $, then we have the following asymptotic relationships:
\begin{equation}
I_{\beta,\alpha}\leq I_{\gamma_{0}}+O\left(  N^{-1}\right)  \leq I\leq
I_{u}=I_{G}+K/2+O\left(  N^{-1}\right) , \label{Thm1A}%
\end{equation}
\begin{equation}
I_{e}=I_{G}+O\left(  N^{-1}\right)  , \label{Ie=Ia}%
\end{equation}
\begin{equation}
I_{\beta,\alpha} =I_{\gamma}+O\left(  N^{-1}\right) ,\label{Ibeta1}%
\end{equation}
where
\begin{equation}
I_{\gamma} =\frac{1}{2}\int_{{{\mathcal{X}}}}p(\mathbf{x})\ln\left(
\det\left(  \frac{\mathbf{G}_{\gamma}(\mathbf{x})}{2\pi}\right)  \right)
d\mathbf{x}+H(X) \label{Igam}%
\end{equation}
and $\gamma_{0}=\beta_{0}\left(  1-\beta_{0}\right)  =1/4$ with $\beta
_{0}=1/2$.
\end{theorem}

\begin{remark}
\label{Remark 0} We see from {Theorems} {\ref{Theorem 1} }and {\ref{Theorem 2}%
} that the true mutual information $I$ and the approximation ${I_{e}}$ both
lie between $I_{\beta_{1},1}$ and $I_{u}$, which implies that their values may
be close to each other. For discrete variable $\mathbf{x}$, {Theorem}%
~{\ref{Theorem 3}} tells us that $I{_{e}}$ and ${I}_{d}$ are asymptotically
equivalent (i.e., their difference vanishes) in the limit of large $N$. For
continuous variable $\mathbf{x}$, {Theorem} {\ref{Theorem 4}} tells us that
$I{_{e}}$ and ${I}_{G}$ are asymptotically equivalent in the limit of large
$N$, which means that ${I}_{e}$ and $I$ are also asymptotically equivalent
because $I_{G}$ and $I$ are known to be asymptotically
equivalent~\citep{Huang(2017-information)} .\qed

\end{remark}

\begin{remark}
\label{Remark 1} To see how the condition in Equation (\ref{Thm1A.1a}) could
be satisfied, consider the case where $D\left(  \mathbf{x}||{\hat{\mathbf{x}}%
}\right)  $ has only one extreme point at ${\hat{\mathbf{x}}}=\mathbf{x}$\ for
${\hat{\mathbf{x}}}\in{{\mathcal{X}}}_{\omega}\left(  {\mathbf{x}}\right)  $
and there exists an $\eta>0$ such that\ $N^{-1}D\left(  \mathbf{x}%
|\mathbf{\hat{x}}\right)  \geq\eta$ for ${\hat{\mathbf{x}}}\in{{\mathcal{\bar
{X}}}}_{\omega}\left(  {\mathbf{x}}\right)  $. Now, the condition in Equation
(\ref{Thm1A.1a}) is satisfied because
\begin{align}
&  \det\left(  \mathbf{G}(\mathbf{x})\right)  ^{1/2}\int_{{{\mathcal{\bar{X}}%
}}_{\varepsilon}\left(  {\mathbf{x}}\right)  }p({\hat{\mathbf{x}}})\exp\left(
-D\left(  \mathbf{x}||{\hat{\mathbf{x}}}\right)  \right)  d\mathbf{\hat{x}%
}\nonumber\\
&  \leq\det\left(  \mathbf{G}(\mathbf{x})\right)  ^{1/2}\int_{{{\mathcal{\bar
{X}}}}_{\varepsilon}\left(  {\mathbf{x}}\right)  }p({\hat{\mathbf{x}}}%
)\exp\left(  -\hat{\eta}\left(  \varepsilon\right)  N\right)  d{\hat
{\mathbf{x}}}\nonumber\\
&  =O\left(  N^{K/2}e^{-\hat{\eta}\left(  \varepsilon\right)  N}\right)  ,
\label{Rem1.4a}%
\end{align}
where by assumption we can find an $\hat{\eta}\left(  \varepsilon\right)  >0$
for any given $\varepsilon\in\left(  0\text{,\thinspace}\omega\right)  $. The
condition in Equation (\ref{Thm1A.1b}) can be satisfied in a similar way. When
$\beta_{0}=1/2$, $\beta_{0}D_{\beta_{0}}\left(  \mathbf{x}||{\hat{\mathbf{x}}%
}\right)  $ is the Bhattacharyya distance \citep{Bhattacharyya(1943-measure)}:
\begin{equation}
\beta_{0}D_{\beta_{0}}\left(  \mathbf{x}||{\hat{\mathbf{x}}}\right)
=-\ln\left(  \int_{{\mathcal{R}}}\sqrt{p(\mathbf{r}|\mathbf{x})p(\mathbf{r}%
|\hat{\mathbf{x}})}d\mathbf{r}\right)  , \label{Rem1.4b}%
\end{equation}
and we have%
\begin{equation}
\mathbf{J}\left(  \mathbf{x}\right)  =\left.  \frac{\partial^{2}\left(
D\left(  \mathbf{x}||{\hat{\mathbf{x}}}\right)  \right)  }{\partial
{\hat{\mathbf{x}}}\partial{\hat{\mathbf{x}}}^{T}}\right\vert _{\hat
{\mathbf{x}}=\mathbf{x}} =\left.  \frac{\partial^{2}\left(  4\beta_{0}%
D_{\beta_{0}}\left(  \mathbf{x}||{\hat{\mathbf{x}}}\right)  \right)
}{\partial{\hat{\mathbf{x}}}\partial{\hat{\mathbf{x}}}^{T}}\right\vert
_{\hat{\mathbf{x}}=\mathbf{x}} =\left.  \frac{\partial^{2}\left(  8H_{l}%
^{2}\left(  \mathbf{x}||{\hat{\mathbf{x}}}\right)  \right)  }{\partial
{\hat{\mathbf{x}}}\partial{\hat{\mathbf{x}}}^{T}}\right\vert _{\hat
{\mathbf{x}}=\mathbf{x}}, \label{Rem1.4c}%
\end{equation}
where $H_{l}\left(  \mathbf{x}||{\hat{\mathbf{x}}}\right)  $ is the Hellinger
distance~\citep{Beran(1977-minimum)}\ between ${p(\mathbf{r}|\mathbf{x})}$ and
${p(\mathbf{r}|{\hat{\mathbf{x}}})}$:
\begin{equation}
H_{l}^{2}\left(  \mathbf{x}||{\hat{\mathbf{x}}}\right)  =\frac{1}{2}%
\int_{{\mathcal{R}}}\left(  \sqrt{p(\mathbf{r}|\mathbf{x})}-\sqrt
{p(\mathbf{r}|{{\hat{\mathbf{x}}}})}\right)  ^{2}d\mathbf{r}. \label{Hel2}%
\end{equation}
By Jensen's inequality,\ for $\beta\in\left(  0\text{,\thinspace}1\right)
$\ we get%
\begin{equation}
0\leq D_{\beta}\left(  \mathbf{x}||{\hat{\mathbf{x}}}\right)  \leq D\left(
\mathbf{x}||{\hat{\mathbf{x}}}\right)  . \label{Rem1.1}%
\end{equation}
Denoting the Chernoff information~\citep{Cover(2006-BK-elements)} as%
\begin{equation}
C\left(  \mathbf{x}||{\hat{\mathbf{x}}}\right)  ={\underset{\beta\in\left(
0\text{,\thinspace}1\right)  }{\max}}\left(  \beta D_{\beta}\left(
\mathbf{x}||{\hat{\mathbf{x}}}\right)  \right)  =\beta_{m}D_{\beta_{m}}\left(
\mathbf{x}||{\hat{\mathbf{x}}}\right)  , \label{Rem1.2}%
\end{equation}
where $\beta D_{\beta}\left(  \mathbf{x}||{\hat{\mathbf{x}}}\right)  $
achieves its maximum at $\beta_{m}$, we have%
\begin{align}
&  I_{\beta,\alpha}-H(X)\nonumber\\
&  \leq h_{c}=-\sum\limits_{m=1}^{M}p\left(  \mathbf{x}_{m}\right)  \ln\left(
\sum\limits_{\hat{m}=1}^{M}\frac{p\left(  \mathbf{x}_{\hat{m}}\right)
}{p\left(  \mathbf{x}_{m}\right)  }\exp\left(  -C\left(  \mathbf{x}%
_{m}||\mathbf{x}_{\hat{m}}\right)  \right)  \right) \label{Rem1.2a}\\
&  \leq h_{d}=-\sum\limits_{m=1}^{M}p\left(  \mathbf{x}_{m}\right)  \ln\left(
\sum\limits_{\hat{m}=1}^{M}\frac{p\left(  \mathbf{x}_{\hat{m}}\right)
}{p\left(  \mathbf{x}_{m}\right)  }\exp\left(  -\beta_{m}D_{\beta}\left(
\mathbf{x}_{m}||\mathbf{x}_{\hat{m}}\right)  \right)  \right)  .
\label{Rem1.2b}%
\end{align}
By {Theorem \ref{Theorem 4}},
\begin{align}
&  \max_{\beta\in\left(  0\text{,\thinspace}1\right)  }I_{\beta,\alpha
}=I_{\gamma_{0}}+O\left(  N^{-1}\right)  ,\label{Rem1.3a}\\
&  I_{\gamma_{0}}=I_{G}-\frac{K}{2}\ln\frac{4}{e}. \label{Rem1.3b}%
\end{align}
If $\beta_{m}=1/2$, then, by Equations (\ref{I->Ie}), (\ref{Rem1.2b}),
(\ref{Rem1.3a}) and (\ref{Rem1.3b}), we have
\begin{align}
\max_{\beta\in\left(  0\text{,\thinspace}1\right)  }I_{\beta}+\frac{K}{2}%
\ln\frac{4}{e}+O\left(  N^{-1}\right)   &  \leq I_{e}={I}+O\left(
N^{-1}\right) \nonumber\\
&  \leq h_{d}+H(X)\leq I_{u}. \label{Rem1.3c}%
\end{align}
Therefore, from Equations (\ref{Rem1.2a}), (\ref{Rem1.2b}) and (\ref{Rem1.3c}%
), we can see that $I_{e}$ and $I$ are close to $h_{c}+H(X)$.\qed

\end{remark}

\subsection{Approximations for Mutual Information}

\label{Sec:2.2a}

In this section, we use the relationships described above to find effective
approximations to true mutual information $I$ in the case of large but finite
$N$. First, {Theorems} {\ref{Theorem 1} }and {\ref{Theorem 2}} tell us that
the true mutual information $I$ and its approximation ${I_{e}}$\ lie between
lower and upper bounds given by:
%$I_{\beta,\alpha}$ and $I_{u}$ or between $I_{\beta
%_{1},1}\ $and $I_{u}$; that is,
$I_{\beta,\alpha}\leq{I}\leq I_{u}$ and $I_{\beta_{1},1}\leq{I_{e}}\leq I_{u}%
$. As a special case, $I$ is also bounded by $I_{\beta_{1},1}\leq{I}\leq
I_{u}$. Furthermore, from Equation (\ref{Ia}) and (\ref{Ie=Ia}) we can obtain
the following asymptotic equality under suitable conditions:
\begin{equation}
{I=I_{e}}+O\left(  N^{-1}\right)  . \label{I->Ie}%
\end{equation}

Hence, for continuous stimuli, we have the following approximate relationship
for large $N$:
\begin{equation}
I\simeq I_{e}\simeq I_{G}. \label{I=Ie}%
\end{equation}

For discrete stimuli, by Equation (\ref{Thm1a.0}) for large but finite $N$, we
have%
\begin{equation}
I\simeq I_{e}\simeq I_{d}=-\sum\limits_{m=1}^{M}p(\mathbf{x}_{m})\ln\left(
1+\sum\limits_{{\hat{m}}\in{\mathcal{M}}_{m}^{u}}\frac{p(\mathbf{x}_{{\hat{m}%
}})}{p(\mathbf{x}_{m})}\exp\left(  -e^{-1}D\left(  \mathbf{x}_{m}%
||\mathbf{x}_{{\hat{m}}}\right)  \right)  \right)  +H\left(  X\right)  .
\label{I=Ied}%
\end{equation}

Consider the special case $p(\mathbf{x}_{{\hat{m}}})\simeq p(\mathbf{x}_{m})$
for ${\hat{m}}\in{\mathcal{M}}_{m}^{u}$. With the help of { Equation }
(\ref{ID}), substitution of $p(\mathbf{x}_{{\hat{m}}})\simeq p(\mathbf{x}%
_{m})$ into Equation (\ref{I=Ied}) yields
\begin{align}
I  &  \simeq I_{D}=-\sum\limits_{m=1}^{M}p(\mathbf{x}_{m})\ln\left(
1+\sum\limits_{{\hat{m}}\in{\mathcal{M}}_{m}^{u}}\exp\left(  -e^{-1}D\left(
\mathbf{x}_{m}||\mathbf{x}_{{\hat{m}}}\right)  \right)  \right)  +H\left(
X\right) \nonumber\\
&  \simeq-\sum\limits_{m=1}^{M}p(\mathbf{x}_{m})\sum\limits_{{\hat{m}}%
\in{\mathcal{M}}_{m}^{u}}\exp\left(  -e^{-1}D\left(  \mathbf{x}_{m}%
||\mathbf{x}_{{\hat{m}}}\right)  \right)  +H\left(  X\right) \nonumber\\
&  =I_{D}^{0} \label{I=Id0}%
\end{align}
where $I_{D}^{0}\leq I_{D}$ and the second approximation follows from the
first-order Taylor expansion assuming that the term $\sum\limits_{{\hat{m}}%
\in{\mathcal{M}}_{m}^{u}}\exp\left(  -e^{-1}D\left(  \mathbf{x}_{m}%
||\mathbf{x}_{{\hat{m}}}\right)  \right)  $ is sufficiently small.

The theoretical discussion above suggests that $I_{e}$\ and $I_{d}$\ are
effective approximations to true mutual information $I$ in the limit of large
$N$. Moreover, we find that they are often good approximations of mutual
information $I$ even for relatively small $N$, as illustrated in the following section.

\section{Results of Numerical Simulations}

Consider Poisson model neuron whose responses (i.e., numbers of spikes within
a given time window) follow a Poisson distribution
\citep{Huang(2017-information)}. The mean response of neuron $n$, with
$n\in\left\{  1\text{,\thinspace}2\text{,\thinspace}\cdots\text{,\thinspace
}N\right\}  $, is described by the tuning function $f\left(
x\text{\textrm{;\thinspace}}\theta_{n}\right)  $, which takes the form of a
Heaviside step function:
\begin{equation}
f\left(  x\text{\textrm{;\thinspace}}\theta_{n}\right)  =\left\{
\begin{array}
[c]{rl}%
A, & \mbox{if $x\geq\theta_{n}$},\\
0, & \mbox{if $x<\theta_{n}$},
\end{array}
\right.  \label{fx}%
\end{equation}
where the stimulus $x\in\left[  -T\text{,\thinspace}T\right]  $ with $T=10$,
$A=10$, and the centers $\theta_{1}$,\thinspace$\theta_{2}$,\thinspace$\cdots
$,\thinspace$\theta_{N}$ of the $N$ neurons are uniformly spaced in interval
$\left[  -T\text{,\thinspace}T\right]  $, namely, $\theta_{n}=\left(
n-1\right)  d-T$ with $d=2T/(N-1)\ $ for $N\geq2$, and $\theta_{n}=0$ for
$N=1$. We suppose that the discrete stimulus $x$ has $M=21$ possible values
that are evenly spaced from $-T$ to $T$, namely, $x\in{{\mathcal{X}}}=\left\{
x_{m}:x_{m}=2\left(  m-1\right)  T/(M-1)-T,m=1,2,\cdots,M\right\}  $. Now, the
Kullback--Leibler divergence can be written as
\begin{equation}
D\left(  x_{m}||x_{{\hat{m}}}\right)  =f\left(  x_{m}\text{\textrm{;\thinspace
}}\theta_{n}\right)  \log\left(  \frac{f\left(  x_{m}\text{\textrm{;\thinspace
}}\theta_{n}\right)  }{f\left(  x_{{\hat{m}}}\text{\textrm{;\thinspace}}%
\theta_{n}\right)  }\right)  +f\left(  x_{{\hat{m}}}\text{\textrm{;\thinspace
}}\theta_{n}\right)  -f\left(  x_{m}\text{\textrm{;\thinspace}}\theta
_{n}\right)  . \label{Dp}%
\end{equation}

Thus, we have $\exp\left(  -e^{-1}D\left(  x_{m}||x_{{\hat{m}}}\right)
\right)  =1$ when $f\left(  x_{m}\text{\textrm{;\thinspace}}\theta_{n}\right)
=f\left(  x_{{\hat{m}}}\text{\textrm{;\thinspace}}\theta_{n}\right)  $,
$\exp\left(  -e^{-1}D\left(  x_{m}||x_{{\hat{m}}}\right)  \right)
=\exp\left(  -e^{-1}A\right)  $ when $f\left(  x_{m}\text{\textrm{;\thinspace
}}\theta_{n}\right)  =0$ and $f\left(  x_{{\hat{m}}}\text{\textrm{;\thinspace
}}\theta_{n}\right)  =A$, and $\exp\left(  -e^{-1}D\left(  x_{m}||x_{{\hat{m}%
}}\right)  \right)  =0$ when $f\left(  x_{m}\text{\textrm{;\thinspace}}%
\theta_{n}\right)  =A$ and$f\left(  x_{{\hat{m}}}\text{\textrm{;\thinspace}%
}\theta_{n}\right)  =0$. Therefore, in this case, we have
\begin{equation}
I_{e}=I_{d}. \label{Ie=Id}%
\end{equation}

More generally, this equality holds true whenever the tuning function has
binary values.

\begin{figure}[ptb]
\caption{A comparison of approximations $I_{e}$, $I_{d}$ and $I_{D}$ against
$I_{MC}$ obtained by Monte Carlo method for one-dimensional discrete stimuli.
(\textbf{a}) Discrete uniform distribution of the stimulus $p(x)$ (black dots)
and the Heaviside step tuning function $f\left(  x\text{\textrm{;\thinspace}%
}\theta\right)  $ with center $\theta=0$ (blue dashed lines); (\textbf{b}) The
values of $I_{MC}$, $I_{e}$, $I_{d}$ and $I_{D}$ depend on the population size
or total number of neurons $N${;} (\textbf{c}) The relative errors $DI_{e}$,
$DI_{d}$ and $DI_{D}$ for the results in ({b}){;} (\textbf{d}) The absolute
values of the relative errors $|DI_{e}|$, $|DI_{d}|$ and $|DI_{D}|$ as in
({c}), with error bars showing standard deviations of repeated trials.}%
\label{Fig1}
%\renewcommand{\baselinestretch}{1.0} \hfill
%\par
\centering\subfigure{\label{Fig1a}
\includegraphics[width= .48\columnwidth]{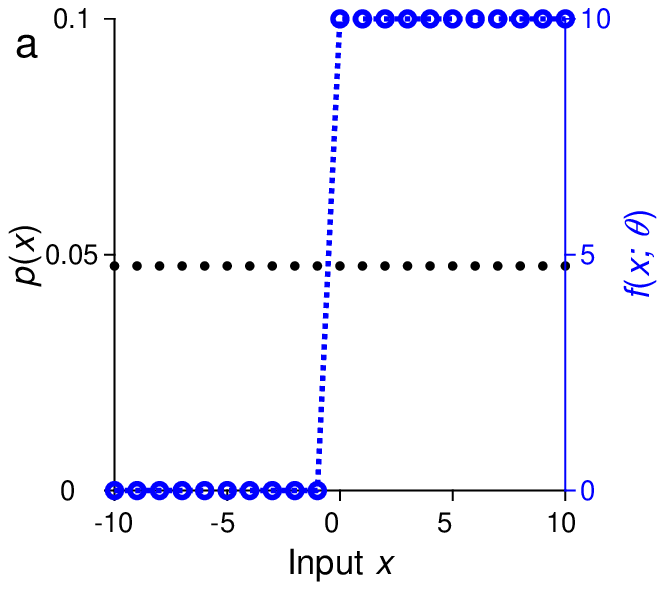}}\hspace{0pt}
\subfigure{\label{Fig1b}
\includegraphics[width= .48\columnwidth]{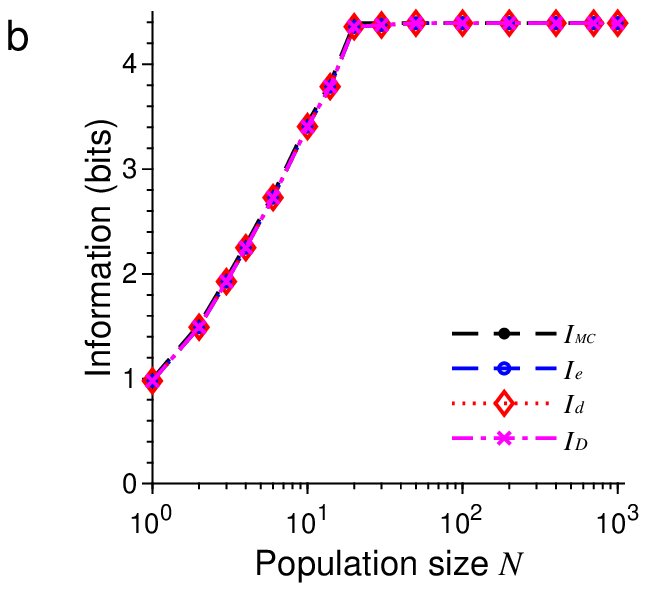}}\hspace{0pt}
\subfigure{\label{Fig1c}
\includegraphics[width= .48\columnwidth]{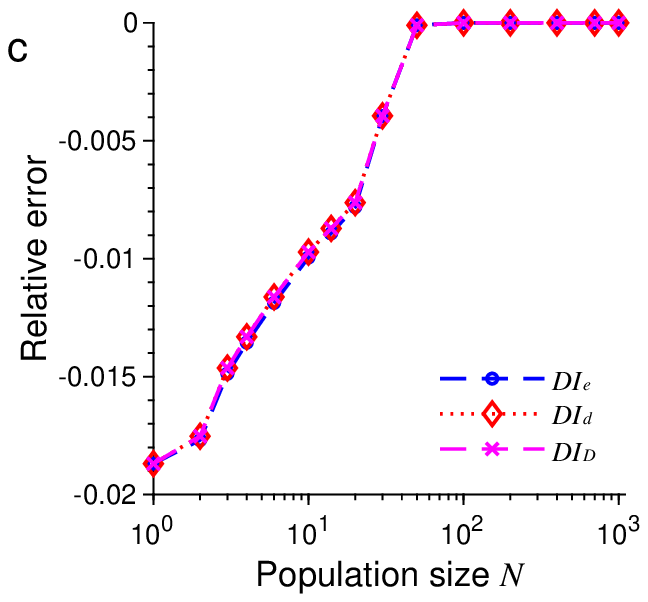}}\hspace{0pt}
\subfigure{\label{Fig1d}
\includegraphics[width= .48\columnwidth]{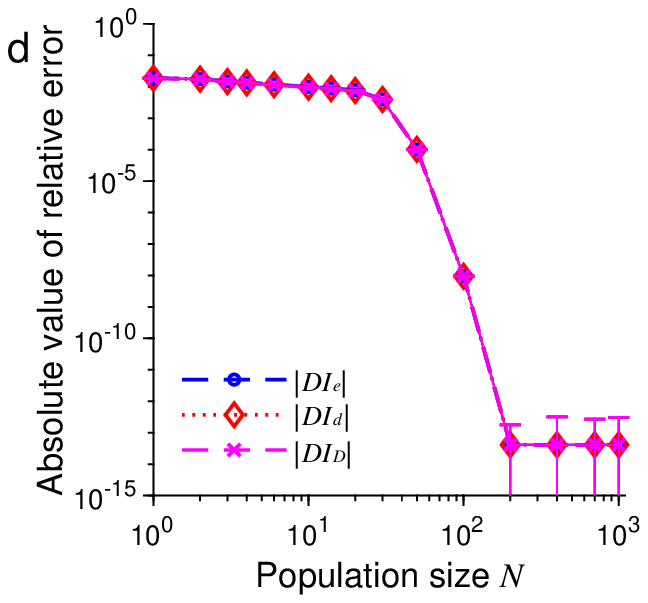}}\hspace{0pt}
\vskip-0.2 in
\par
\vspace{18pt}
\par
For the second example, we only changed the probability distribution of
stimulus $p(x_{m})$ while keeping all other conditions unchanged. Now,
$p(x_{m})$ is a discrete sample from a Gaussian function:
\begin{equation}
p(x_{m})=Z^{-1}\exp\left(  -\frac{x_{m}^{2}}{2\sigma^{2}}\right)
,m=1,2,\cdots,M, \label{px}%
\end{equation}
where $Z$ is the normalization constant and $\sigma=T/2$. The results are
illustrated in {Figure~\ref{Fig2}}.\end{figure}

In the first example, as illustrated in {Figure~\ref{Fig1}}, we suppose the
stimulus has a uniform distribution, so that the probability is given by
$p(x_{m})=1/M$. {Figure~\ref{Fig1}a} shows graphs of the input distribution
$p(x)$ and a representative tuning function $f\left(
x\text{\textrm{;\thinspace}}\theta\right)  $ with the center $\theta=0$.

To assess the accuracy of the approximation formulas, we employed Monte Carlo
(MC) simulation to evaluate the mutual information $I$
\citep{Huang(2017-information)}. In our MC simulation, we first sampled an
input $x_{j}\in{{\mathcal{X}}}$ from the uniform distribution $p(x_{j})=1/M$,
then generated the neural responses $\mathbf{r}_{j}$ by the conditional
distribution $p(\mathbf{r}_{j}|x_{j})$ based on the Poisson model,
where\ $j=1$,\thinspace$2$,\thinspace$\cdots$,\thinspace$j_{\mathrm{\max}}$.
The value of mutual information by MC simulation was calculated by
\begin{equation}
I_{MC}^{\ast}=\frac{1}{j_{\mathrm{\max}}}\sum\limits_{j=1}^{j_{\mathrm{\max}}%
}\ln\left(  \frac{p(\mathbf{r}_{j}|x_{j})}{p(\mathbf{r}_{j})}\right)  ,
\label{IMC*}%
\end{equation}
where
\begin{equation}
p(\mathbf{r}_{j})=\sum\limits_{m=1}^{M}p(\mathbf{r}_{j}|x_{m})p(x_{m}).
\label{prj}%
\end{equation}

To assess the precision of our MC simulation, we computed the standard
deviation of repeated trials by bootstrapping:
\begin{equation}
I_{std}=\sqrt{\frac{1}{i_{\mathrm{\max}}}\sum\limits_{i=1}^{i_{\mathrm{\max}}%
}\left(  I_{MC}^{i}-I_{MC}\right)  ^{2}}, \label{Istd}%
\end{equation}
where
\begin{align}
I_{MC}^{i}  &  =\frac{1}{j_{\mathrm{\max}}}\sum\limits_{j=1}^{j_{\mathrm{\max
}}}\ln\left(  \frac{p(\mathbf{r}_{\Gamma_{j,i}}|x_{\Gamma_{j,i}}%
)}{p(\mathbf{r}_{\Gamma_{j,i}})}\right)  ,\label{IMCi}\\
I_{MC}  &  =\frac{1}{i_{\mathrm{\max}}}\sum\limits_{i=1}^{i_{\mathrm{\max}}%
}I_{MC}^{i}, \label{IMC}%
\end{align}
and $\Gamma_{j,i}\in\left\{  1\text{,\thinspace}2\text{,\thinspace}%
\cdots\text{,\thinspace}j_{\mathrm{\max}}\right\}  $ is the $\left(
j,i\right)  $-th entry of the matrix $\boldsymbol{\Gamma}\in\mathcal{%
%TCIMACRO{\U{2115} }%
%BeginExpansion
\mathbb{N}
%EndExpansion
}^{j_{\mathrm{\max}}\times i_{\mathrm{\max}}}$ with samples taken randomly
from the integer set $\{1$,\thinspace$2$,\thinspace$\cdots$,\thinspace
$j_{\mathrm{\max}}\}$ by a uniform distribution. Here, we set $j_{\mathrm{\max
}}=5\times10^{5}$, $i_{\mathrm{\max}}=100$ and $M=10^{3}$.

For different $N\in\{1$,$\,2$,$\,3$,$\,4$,$\,6$,$\,10$,$\,14$,$\,20$%
,$\,30$,$\,50$,$\,100$,$\,200$,$\,400$,$\,700$,$\,1000\}$, we compared
$I_{MC}$\ with $I_{e}$, $I_{d}$ and $I_{D}$, as illustrated in {Figure
\ref{Fig1}b--d}. Here, we define the relative error of approximation, e.g.,
for $I_{e}$, as
\begin{equation}
DI_{e}=\frac{I_{e}-I_{MC}}{I_{MC}}, \label{RLError}%
\end{equation}
and the relative standard deviation
\begin{equation}
DI_{std}=\frac{I_{std}}{I_{MC}}. \label{DIstd}%
\end{equation}

\begin{figure}[ptb]
%\renewcommand{\baselinestretch}{1.0} \hfill
%\par
\centering\subfigure{\label{Fig2a}
\includegraphics[width= .48\columnwidth]{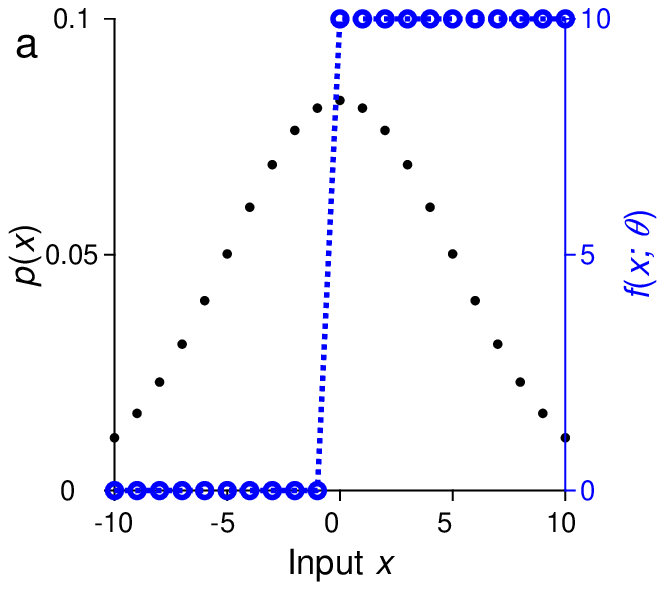}}\hspace{0pt}
\subfigure{\label{Fig2b}
\includegraphics[width= .48\columnwidth]{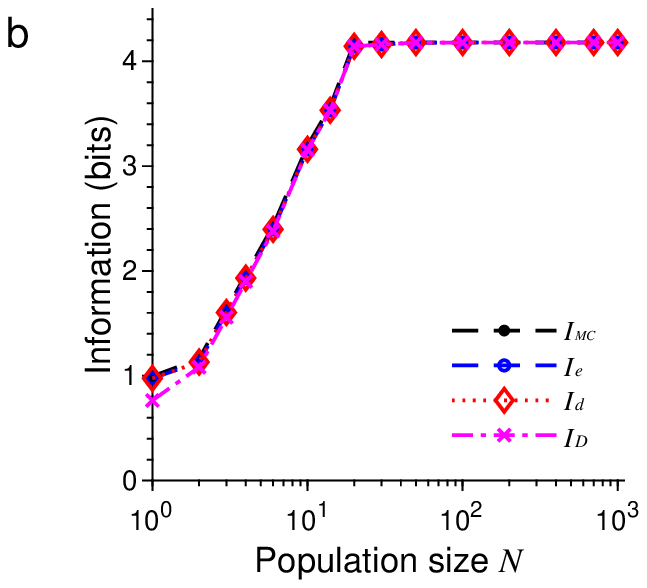}}\hspace{0pt}
\subfigure{\label{Fig2c}
\includegraphics[width= .48\columnwidth]{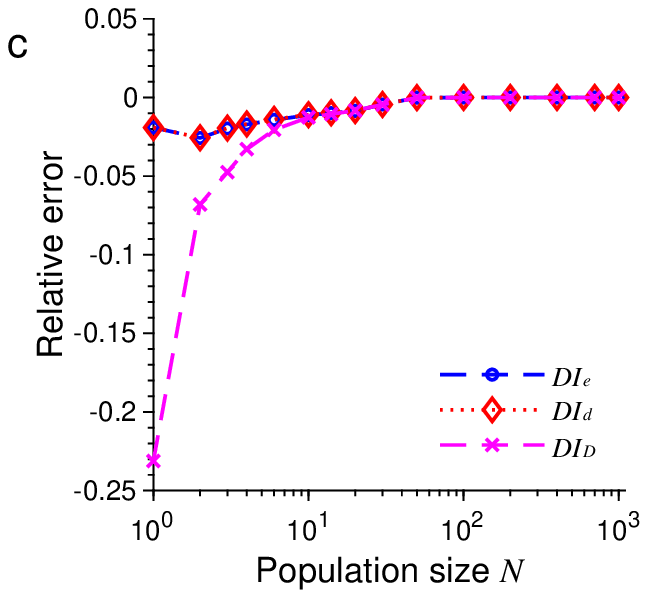}}\hspace{0pt}
\subfigure{\label{Fig2d}
\includegraphics[width= .48\columnwidth]{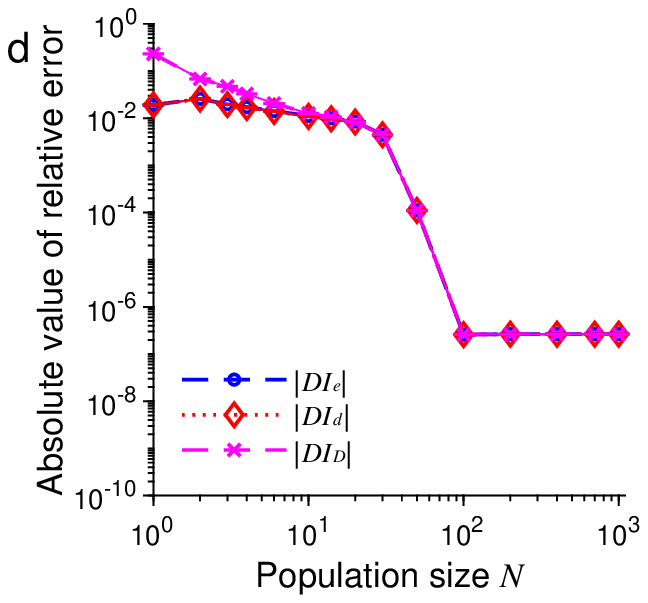}}\hspace{0pt}
\vskip-0.2 in
%There is no explanation for subfigures
\caption{A comparison of approximations $I_{e}$, $I_{d}$ and $I_{D}$ against
$I_{MC}$. The situation is identical to that in Figure~\ref{Fig1} except that
the stimulus distribution $p(x)$ is peaked rather flat (black dots in ({a})).
(\textbf{a}) {Discrete Gaussian-like distribution of the stimulus $p(x)$
(black dots) and the Heaviside step tuning function $f\left(
x\text{\textrm{;\thinspace}}\theta\right)  $ with center $\theta=0$ (blue
dashed lines);} (\textbf{b}) {The values of $I_{MC}$, $I_{e}$, $I_{d}$ and
$I_{D}$ depend on the population size or total number of neurons $N$;}
(\textbf{c}) {The relative errors $DI_{e}$, $DI_{d}$ and $DI_{D}$ for the
results in ({b});} (\textbf{d}) {The absolute values of the relative errors
$|DI_{e}|$, $|DI_{d}|$ and $|DI_{D}|$ as in ({c}), with error bars showing
standard deviations of repeated trials}.}%
\label{Fig2}%
\end{figure}

Next, we changed each tuning function $f\left(  x\text{\textrm{;\thinspace}%
}\theta_{n}\right)  $ to a rectified linear function:%
\begin{equation}
f\left(  x;\theta_{n}\right)  =\max\left(  0,x-\theta_{n}\right)  ,
\label{ReLU}%
\end{equation}

\begin{figure}[ptb]
%\renewcommand{\baselinestretch}{1.0} \hfill
%\par
\centering\subfigure{\label{Fig3a}
\includegraphics[width= .48\columnwidth]{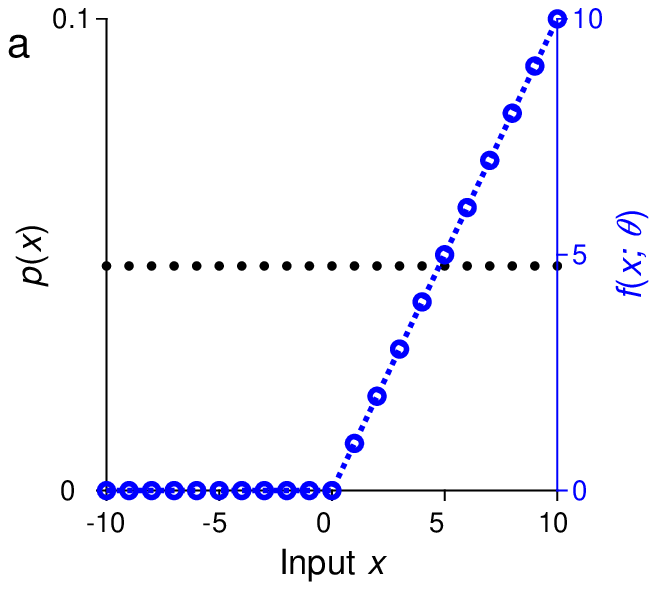}}\hspace{0pt}
\subfigure{\label{Fig3b}
\includegraphics[width= .48\columnwidth]{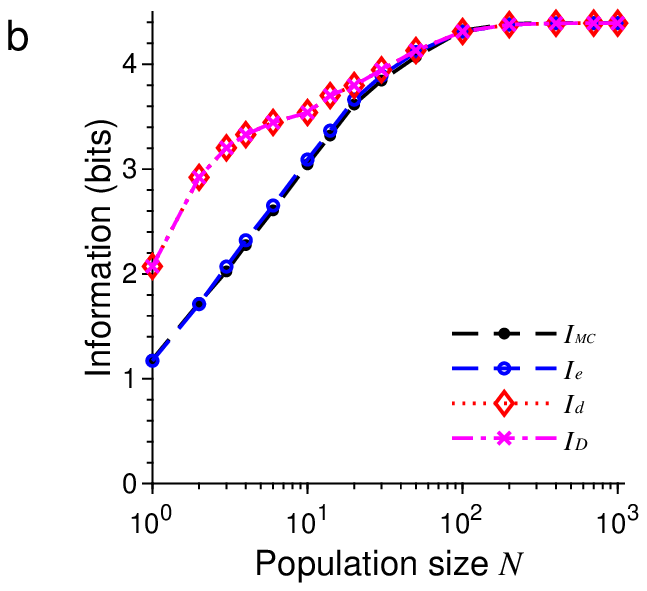}}\hspace{0pt}
\subfigure{\label{Fig3c}
\includegraphics[width= .48\columnwidth]{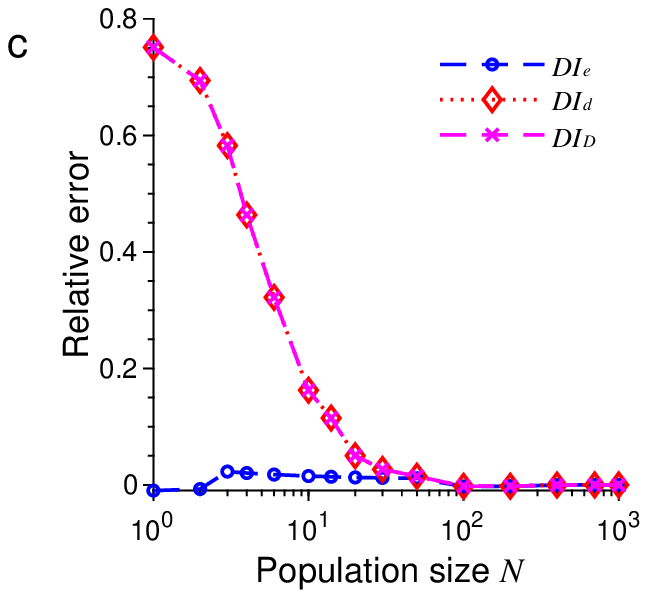}}\hspace{0pt}
\subfigure{\label{Fig3d}
\includegraphics[width= .48\columnwidth]{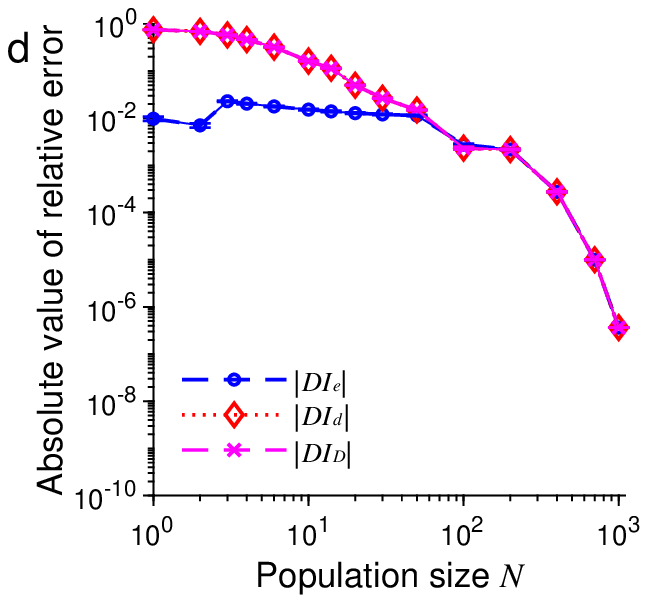}}\hspace{0pt}
\vskip-0.2 in
%There is no explanation for subfigures
\caption{A comparison of approximations $I_{e}$, $I_{d}$ and $I_{D}$ against
$I_{MC}$. The situation is identical to that in Figure~\ref{Fig1} except for
the shape of the tuning function (blue dashed lines in (a)). (\textbf{a})
{Discrete uniform distribution of the stimulus $p(x)$ (black dots) and the
rectified linear tuning function $f\left(  x\text{\textrm{;\thinspace}}%
\theta\right)  $ with center $\theta=0$ (blue dashed lines)}; (\textbf{b})
{The values of $I_{MC}$, $I_{e}$, $I_{d}$ and $I_{D}$ depend on the population
size or total number of neurons $N$}; (\textbf{c}) {The relative errors
$DI_{e}$, $DI_{d}$ and $DI_{D}$ for the results in ({b})}; (\textbf{d}) {The
absolute values of the relative errors $|DI_{e}|$, $|DI_{d}|$ and $|DI_{D}|$
as in ({c}), with error bars showing standard deviations of repeated trials}.
}%
\label{Fig3}%
\end{figure}

\begin{figure}[ptb]
%\renewcommand{\baselinestretch}{1.0} \hfill
%\par
\centering\subfigure{\label{Fig4a}
\includegraphics[width= .48\columnwidth]{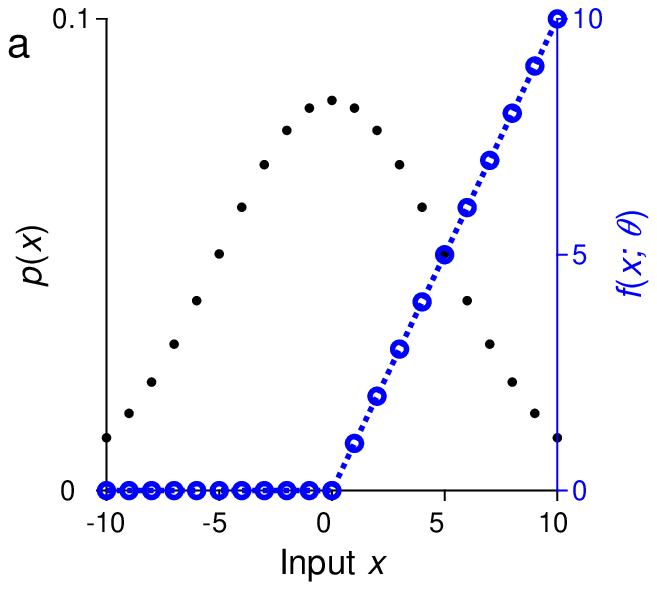}}\hspace{0pt}
\subfigure{\label{Fig4b}
\includegraphics[width= .48\columnwidth]{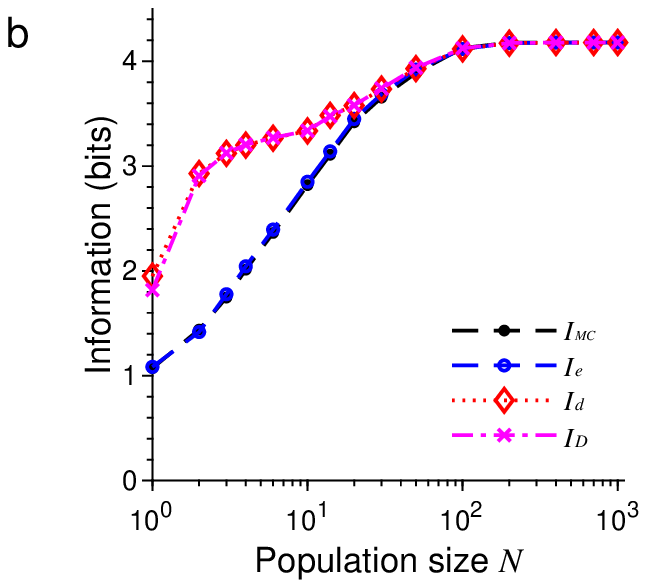}}\hspace{0pt}
\subfigure{\label{Fig4c}
\includegraphics[width= .48\columnwidth]{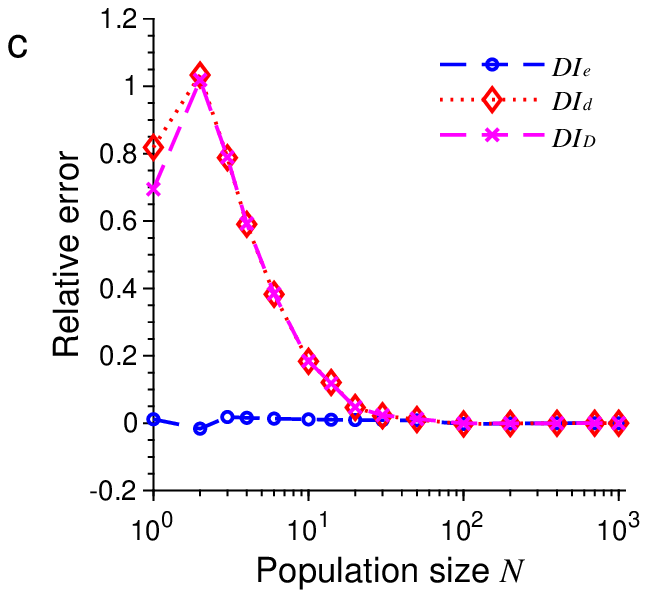}}\hspace{0pt}
\subfigure{\label{Fig4d}
\includegraphics[width= .48\columnwidth]{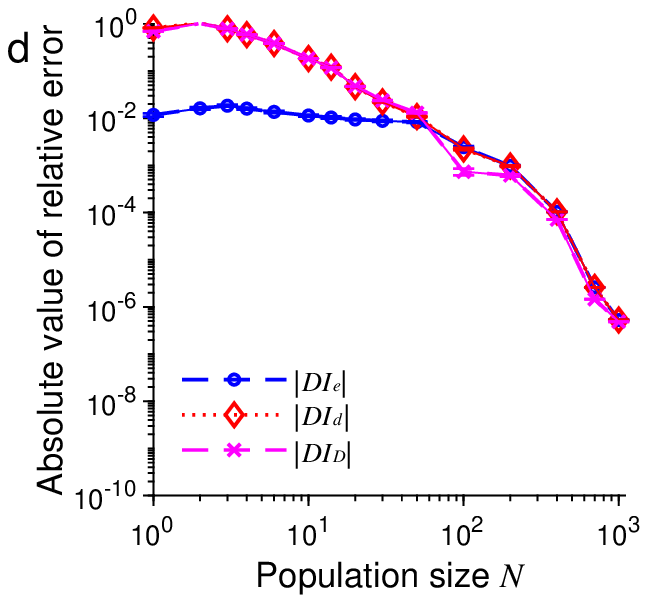}}\hspace{0pt}
\vskip-0.2 in
%%There is no explanation for subfigures
\caption{A comparison of approximations $I_{e}$, $I_{d}$ and $I_{D}$ against
$I_{MC}$. The situation is identical to that in Figure~\ref{Fig3} except that
the stimulus distribution $p(x)$ is peaked rather flat (black dots in (a)).
(\textbf{a}) {Discrete Gaussian-like distribution of the stimulus $p(x)$
(black dots) and the rectified linear tuning function $f\left(
x\text{\textrm{;\thinspace}}\theta\right)  $ with center $\theta=0$ (blue
dashed lines)}; (\textbf{b}) {The values of $I_{MC}$, $I_{e}$, $I_{d}$ and
$I_{D}$ depend on the population size or total number of neurons $N$};
(\textbf{c}) {The relative errors $DI_{e}$, $DI_{d}$ and $DI_{D}$ for the
results in ({b})}; (\textbf{d}) {The absolute values of the relative errors
$|DI_{e}|$, $|DI_{d}|$ and $|DI_{D}|$ as in ({c}), with error bars showing
standard deviations of repeated trials}. }%
\label{Fig4}%
\end{figure}

{Figures \ref{Fig3}} and {\ref{Fig4}} show the results under the same
conditions of Figures \ref{Fig1} and \ref{Fig2} except for the shape of the
tuning functions.

\begin{figure}[ptb]
%\renewcommand{\baselinestretch}{1.0} \hfill
%\par
\centering\subfigure{\label{Fig5a}
\includegraphics[width= .48\columnwidth]{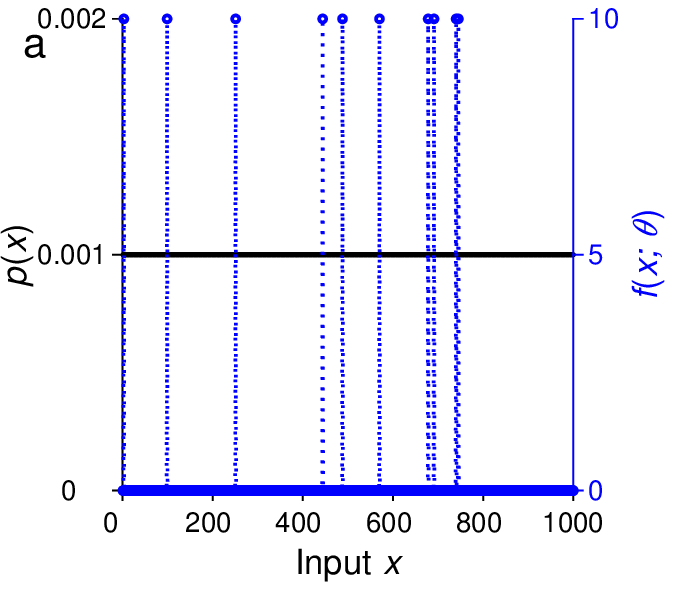}}\hspace{0pt}
\subfigure{\label{Fig5b}
\includegraphics[width= .48\columnwidth]{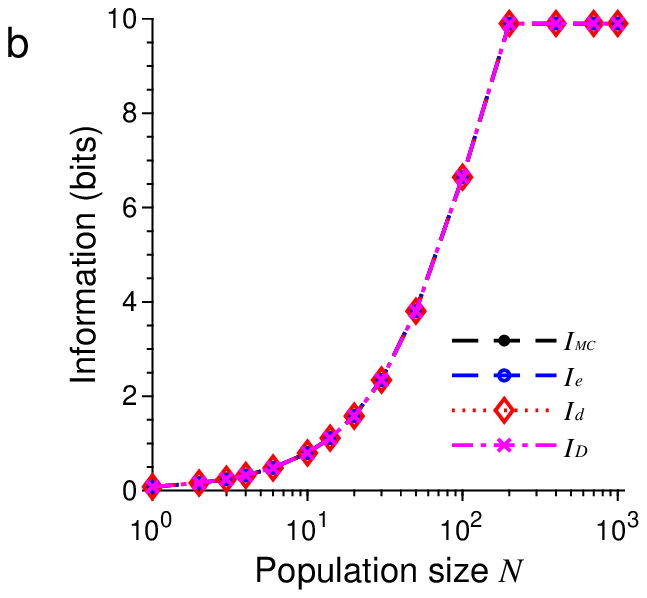}}\hspace{0pt}
\subfigure{\label{Fig5c}
\includegraphics[width= .48\columnwidth]{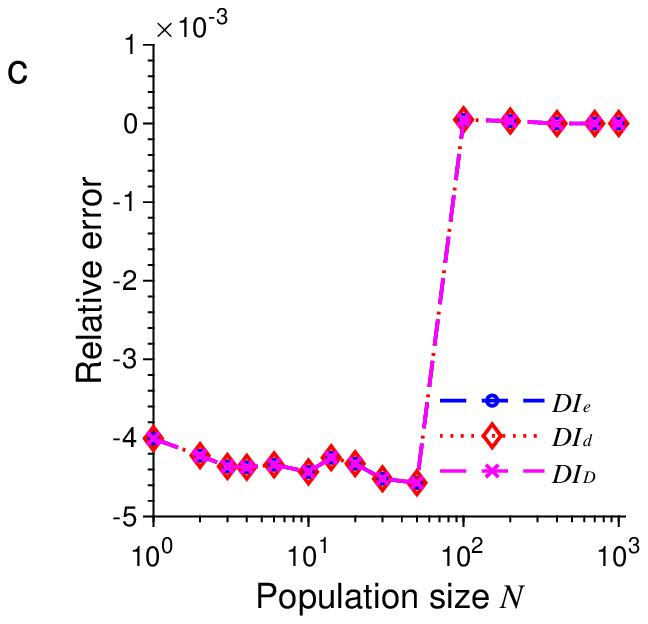}}\hspace{0pt}
\subfigure{\label{Fig5d}
\includegraphics[width= .48\columnwidth]{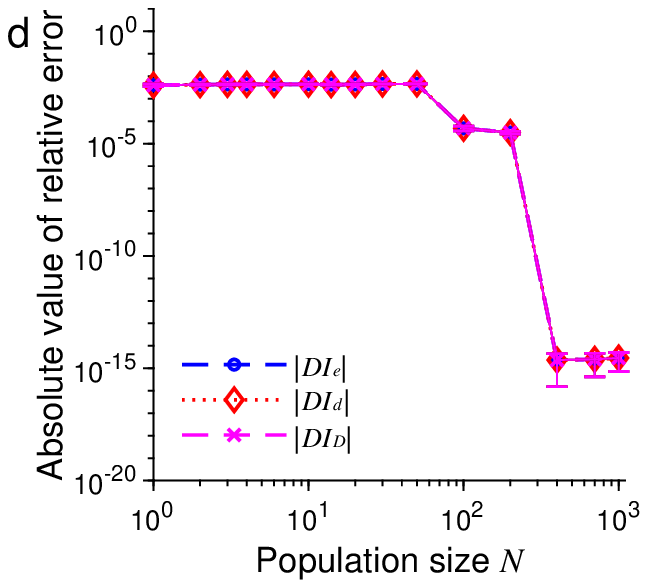}}\hspace{0pt}
\vskip-0.2 in\caption{A comparison of approximations $I_{e}$, $I_{d}$ and
$I_{D}$ against $I_{MC}$. The situation is similar to that in
Figure~\ref{Fig1} except that the tuning function is random (blue dashed lines
in (a)); see Equation~\eqref{fx1}. (\textbf{a}) {Discrete uniform distribution
of the stimulus $p(x)$ (black dots) and the random tuning function $f\left(
x\text{\textrm{;\thinspace}}\theta\right)  $}; (\textbf{b}) {The values of
$I_{MC}$, $I_{e}$, $I_{d}$ and $I_{D}$ depend on the population size or total
number of neurons $N$}; (\textbf{c}) {The relative errors $DI_{e}$, $DI_{d}$
and $DI_{D}$ for the results in ({b})}; (\textbf{d}) {The absolute values of
the relative errors $|DI_{e}|$, $|DI_{d}|$ and $|DI_{D}|$ as in ({c}), with
error bars showing standard deviations of repeated trials}.}%
\label{Fig5}%
\end{figure}

\begin{figure}[ptb]
\centering\subfigure{\label{Fig6a}
\includegraphics[width= .48\columnwidth]{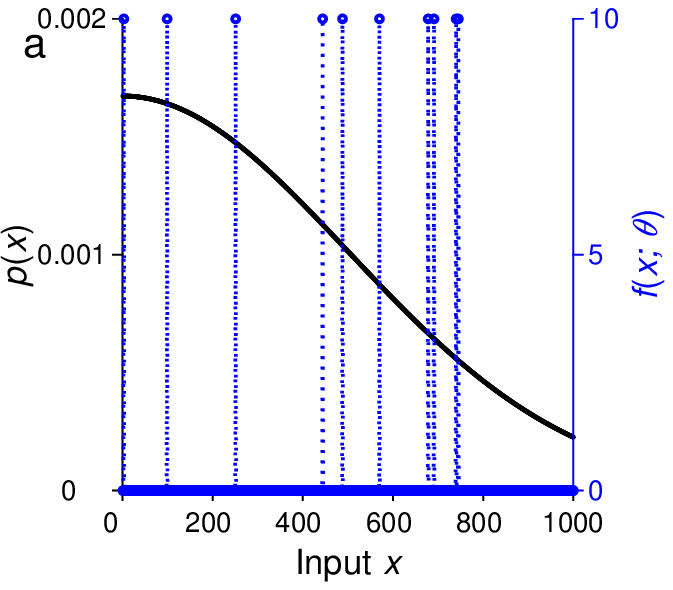}}\hspace{0pt}
\subfigure{\label{Fig6b}
\includegraphics[width= .48\columnwidth]{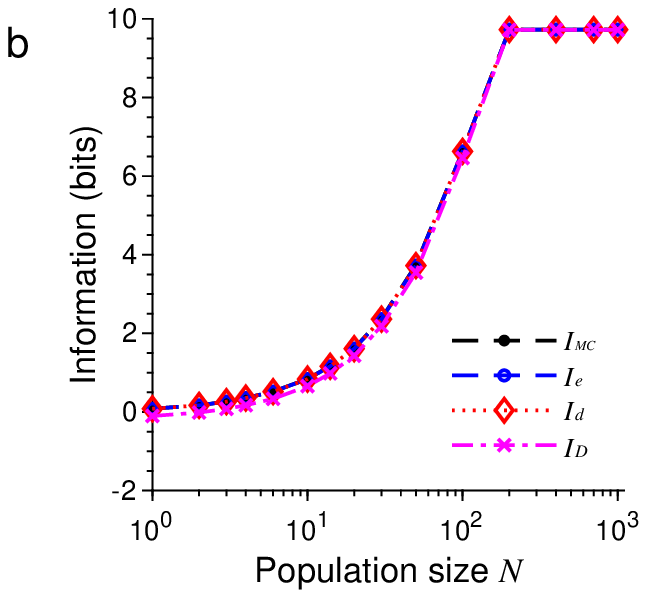}}\hspace{0pt}
\subfigure{\label{Fig6c}
\includegraphics[width= .48\columnwidth]{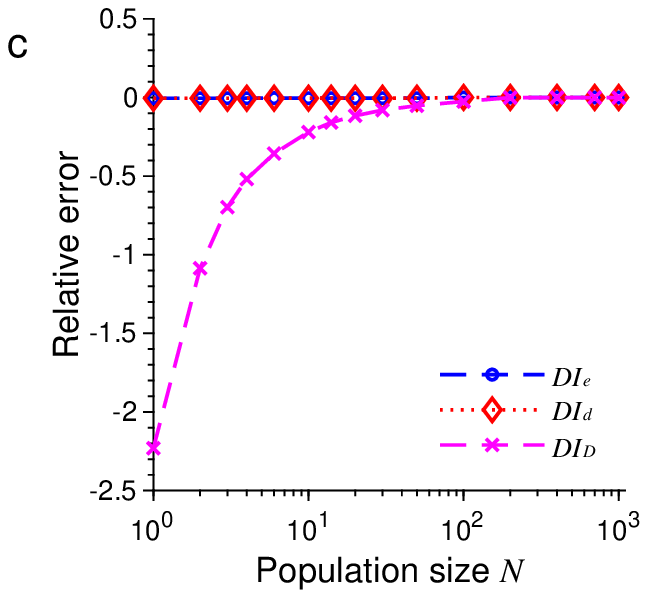}}\hspace{0pt}
\subfigure{\label{Fig6d}
\includegraphics[width= .48\columnwidth]{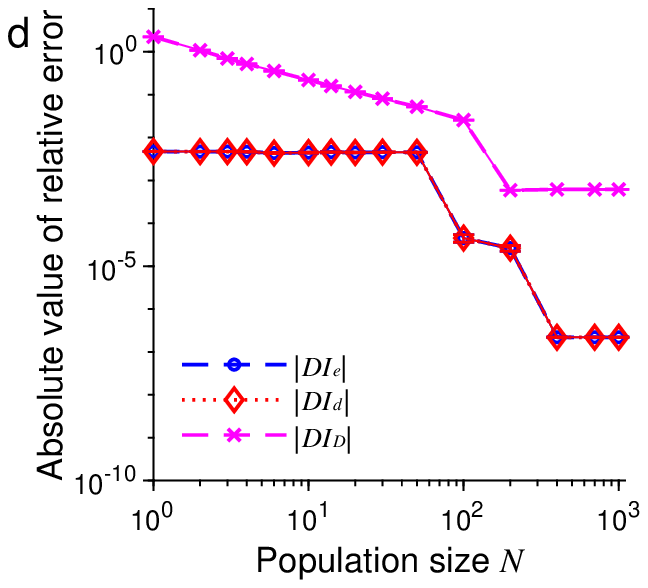}}\hspace{0pt}
\vskip-0.2 in
%There is no explanation for subfigures
\caption{A comparison of approximations $I_{e}$, $I_{d}$ and $I_{D}$ against
$I_{MC}$. The situation is identical to that in Figure~\ref{Fig5} except that
the stimulus distribution $p(x)$ is not flat (black dots in (a)). (\textbf{a})
{Discrete Gaussian-like distribution of the stimulus $p(x)$ (black dots) and
the random tuning function $f\left(  x\text{\textrm{;\thinspace}}%
\theta\right)  $}; (\textbf{b}) {The values of $I_{MC}$, $I_{e}$, $I_{d}$ and
$I_{D}$ depend on the population size or total number of neurons $N$};
(\textbf{c}) {The relative errors $DI_{e}$, $DI_{d}$ and $DI_{D}$ for the
results in ({b})}; (\textbf{d}) {The absolute values of the relative errors
$|DI_{e}|$, $|DI_{d}|$ and $|DI_{D}|$ as in ({c}), with error bars showing
standard deviations of repeated trials}. }%
\label{Fig6}%
\end{figure}

Finally, we let the tuning function $f\left(  x\text{\textrm{;\thinspace}%
}\theta_{n}\right)  $ have a random form:
\begin{equation}
f\left(  x\text{\textrm{;\thinspace}}\theta_{n}\right)  =\left\{
\begin{array}
[c]{rl}%
B, & \mbox{if $\ x\in\theta_{n}=\left\{  \theta
_{n}^{1}\text{,\thinspace}\theta_{n}^{2}\text{,\thinspace}\cdots
\text{,\thinspace}\theta_{n}^{K}\right\}  $},\\
0, & \mathrm{otherwise},
\end{array}
\right.  \label{fx1}%
\end{equation}
where the stimulus $x\in\mathcal{X}=\{1$,\thinspace$2$,\thinspace$\cdots
$,\thinspace$999$,\thinspace$1000\}$, $B=10$, the values of $\left\{
\theta_{n}^{1}\text{,\thinspace}\theta_{n}^{2}\text{,\thinspace}%
\cdots\text{,\thinspace}\theta_{n}^{K}\right\}  $ are distinct and randomly
selected from the set $\mathcal{X}$ with $K=10$. In this example, we may
regard $\mathcal{X}$ as a list of natural objects (stimuli), and there are a
total of $N$ sensory neurons, each of which responds only to $K$ randomly
selected objects. {Figure \ref{Fig5}} shows the results under the condition
that $p(x)$ is a uniform distribution. In {Figure \ref{Fig6}}, we assume that
$p(x)$ is not flat but a half Gaussian given by Equation (\ref{px}) with
$\sigma=500$. 

In all these examples, we found that the three formulas, namely, $I_{e}$,
$I_{d}$ and $I_{D}$, provided excellent approximations to the true values of
mutual information as evaluated by Monte Carlo method. For example, in the
examples in Figures~\ref{Fig1} and ~\ref{Fig5}, all three approximations were
practically indistinguishable. In general, all these approximations were
extremely accurate when $N>100$.

In all our simulations, the mutual information tended to increase with the
population size $N$, eventually reaching a plateau for large enough $N$. The
saturation of information for large $N$ is due to the fact that it requires at
most $\log_{2}M$ bits of information to completely distinguish all $M$
stimuli. It is impossible to gain more information than this maximum amount
regardless of how many neurons are used in the population. In
Figure~\ref{Fig1}, for instance, this maximum is $\log_{2} 21 = 4.39$ bits,
and in Figure~\ref{Fig5}, this maximum is $\log_{2}1000=9.97$ bits.

%There is no explanation for subfigures
%\renewcommand{\baselinestretch}{1.0} \hfill
%\par

For relatively small values of $N$, we found that $I_{D}$ tended to be less
accurate than $I_{e}$ or $I_{d}$ (see Figures \ref{Fig5} and \ref{Fig6}). Our
simulations also confirmed two analytical results. The first one is that
$I_{d}=I_{D}$ when the stimulus distribution is uniform; this result follows
directly from the definitions of $I_{d}$ and $I_{D}$ and is confirmed by the
simulations in Figures \ref{Fig1}, \ref{Fig3}, and \ref{Fig5}. The second
result is that $I_{d}=I_{e}$ (Equation~\eqref{Ie=Id}) when the tuning function
is binary, as confirmed by the simulations in Figures \ref{Fig1}, \ref{Fig2},
\ref{Fig5}, and \ref{Fig6}. When the tuning function allows many different
values, $I_{e}$ can be much more accurate than $I_{d}$ and $I_{D}$, as shown
by the simulations in Figures \ref{Fig3} and \ref{Fig4}. To summarize, our
best approximation formula is $I_{e}$ because it is more accurate than $I_{d}$
and $I_{D}$, and, unlike $I_{d}$ and $I_{D}$, it applies to both discrete and
continuous stimuli (Equations~\eqref{Ie} and \eqref{Rem1.Ie}).

\section{Discussion}

\label{Conclusion}

We have derived several asymptotic bounds and effective approximations of
mutual information for discrete variables and established several
relationships among different approximations. Our final approximation formulas
involve only Kullback--Leibler divergence, which is often easier to evaluate
than Shannon mutual information in practical applications. Although in this
paper our theory is developed in the framework of neural population coding
with concrete examples, our mathematical results are generic and should hold
true in many related situations beyond the original context.

We propose to approximate the mutual information with several asymptotic
formulas, including $I_{e}$ in Equation~\eqref{Ie} or Equation
\eqref{Rem1.Ie}, $I_{d}$ in Equation~\eqref{Id} and $I_{D}$ in
Equation~\eqref{ID}. Our numerical experimental results show that the three
approximations $I_{e}$, $I_{d}$ and $I_{D}$ were very accurate for large
population size $N$, and sometimes even for relatively small $N$. Among the
three approximations, $I_{D}$ tended to be the least accurate, although, as a
special case of $I_{d}$, it is slightly easier to evaluate than $I_{d}$. For a
comparison of $I_{e}$ and $I_{d}$, we note that $I_{e}$ is the universal
formula, whereas $I_{d}$ is restricted only to discrete variables. The two
formulas $I_{e}$ and $I_{d}$ become identical when the responses or the tuning
functions have only two values. For more general tuning functions, the
performance of $I_{e}$ was better than $I_{d}$ in our simulations.

As mentioned before, an advantage of of $I_{e}$ is that it works not only for
discrete stimuli but also for continuous stimuli. Theoretically speaking, the
formula for $I_{e}$ is well justified, and we have proven that it approaches
the true mutual information $I$ in the limit of large population. In our
numerical simulations, the performance of $I_{e}$ was excellent and better
than that of $I_{d}$ and $I_{D}$. Overall, $I_{e}$ is our most accurate and
versatile approximation formula, although, in some cases, $I_{d}$ and $I_{D}$
are slightly more convenient to calculate.

The numerical examples considered in this paper were based on an independent
population of neurons whose responses have Poisson statistics. Although such
models are widely used, they are appropriate only if the neural responses can
be well characterized by the spike counts within a fixed time window. To study
the temporal patterns of spike trains, one has to consider more complicated
models. Estimation of mutual information from neural spike trains is a
difficult computational problem~\citep{Strong1998-entropy,Nemenman2004-entropy,
Panzeri2017-correcting, Houghton2019-calculating}. In future work, it would be
interesting to apply the asymptotic formulas such as $I_{e}$ to spike trains
with small time bins each containing either one spike or nothing. A potential
advantage of the asymptotic formula is that it might help reduce the bias
caused by small samples in the calculation of the response marginal
distribution $p(\mathbf{r})=\sum_{\mathbf{x}} p(\mathbf{r|x})p(\mathbf{x})$ or
the response entropy $H(R)$ because here one only needs to calculate the
Kullback--Leibler divergence $D\left(  \mathbf{x}||{\hat{\mathbf{x}}}\right)
$, which may have a smaller estimation error.

Finding effective approximation methods for computing mutual information is a
key step for many practical applications of the information theory. Generally
speaking, Kullback--Leibler divergence (Equation~\eqref{DKL}) is often easier
to evaluate and approximate than either Chernoff information
(Equation~\eqref{Rem1.2}) or Shannon mutual information (Equation~\eqref{MI}).
In situations where this is indeed the case, our approximation formulas are
potentially useful. Besides applications in numerical simulations, the
availability of a set of approximation formulas may also provide helpful
theoretical tools in future analytical studies of information coding and representations.

As mentioned in the Introduction, various methods have been proposed to
approximate the mutual information~\citep{Miller(1955-note),
Carlton(1969-bias), Treves(1995-upward),Victor(2000-asymptotic),
Paninski(2003-estimation),Kraskov(2004-estimating),Khan(2007-relative),Safaai2018-information}%
. In future work, it would be useful to compare different methods rigorously
under identical conditions in order to asses their relative merits. The
approximation formulas developed in this paper are relatively easy to compute
for practical problems. They are especially suitable for analytical purposes;
for example, they could be used explicitly as objective functions for
optimization or learning algorithms. Although the examples used in our
simulations in this paper are parametric, it should be possible to extend the
formulas to nonparametric problem, possibly with help of the copula method to
take advantage of its robustness in nonparametric
estimations~\citep{Safaai2018-information}.

\medskip

\noindent\textbf{Author Contributions:} {W.H. developed and proved the theorems,
programmed the numerical experiments and wrote the manuscript. K.Z. verified
the proofs and revised the manuscript.}

\noindent\textbf{Funding:}{This research was supported  by an NIH grant R01 DC013698.}

\noindent\textbf{Conflicts of Interest:}{The authors declare no conflict of interest.}

%\appendixtitles{yes} \appendixsections{multiple}

\noindent

\appendix
%\numberwithin{equation}{section}
\renewcommand{\theequation}{A.\arabic{equation}} \setcounter{equation}{0} \renewcommand\thesubsection{A.\arabic{subsection}}

\section*{Appendix: The Proofs}

\section*{The Proofs}

\vspace{-6pt}

\subsection{ Proof of Theorem \ref{Theorem 1}}

\label{A.Lma1aP}By Jensen's inequality, we have
\begin{align}
I_{\beta,\alpha}  &  =-\left\langle \ln\left(  \int_{{{\mathcal{X}}}%
}\left\langle \frac{p^{\beta}(\mathbf{r}|{\hat{\mathbf{x}}})p^{\alpha}%
({\hat{\mathbf{x}}})}{p^{\beta}\left(  \mathbf{r}|\mathbf{x}\right)
p^{\alpha}(\mathbf{x})}\right\rangle _{\mathbf{r}|\mathbf{x}}d{\hat
{\mathbf{x}}}\right)  \right\rangle _{\mathbf{x}}+H(X)\nonumber\\
&  \leq-\left\langle \left\langle \ln\left(  \int_{{{\mathcal{X}}}}%
\frac{p^{\beta}(\mathbf{r}|{\hat{\mathbf{x}}})p^{\alpha}({\hat{\mathbf{x}}}%
)}{p^{\beta}\left(  \mathbf{r}|\mathbf{x}\right)  p^{\alpha}(\mathbf{x}%
)}d{\hat{\mathbf{x}}}\right)  \right\rangle _{\mathbf{r}|\mathbf{x}%
}\right\rangle _{\mathbf{x}}+H(X) \label{A.Lma1aP.a}%
\end{align}
and
\begin{align}
&  -\left\langle \left\langle \ln\left(  \int_{{{\mathcal{X}}}}\frac{p^{\beta
}(\mathbf{r}|{\hat{\mathbf{x}}})p^{\alpha}({\hat{\mathbf{x}}})}{p^{\beta
}\left(  \mathbf{r}|\mathbf{x}\right)  p^{\alpha}(\mathbf{x})}d{\hat
{\mathbf{x}}}\right)  \right\rangle _{\mathbf{r}|\mathbf{x}}\right\rangle
_{\mathbf{x}}+H(X)-I\nonumber\\
&  =\left\langle \left\langle \ln\left(  \int_{{{\mathcal{X}}}}\frac
{p(\mathbf{r},{\hat{\mathbf{x}}})}{p\left(  \mathbf{r},\mathbf{x}\right)
}d{\hat{\mathbf{x}}}\right)  \left(  \int_{{{\mathcal{X}}}}\frac{p^{\beta
}(\mathbf{r}|{\hat{\mathbf{x}}})p^{\alpha}({\hat{\mathbf{x}}})}{p^{\beta
}\left(  \mathbf{r}|\mathbf{x}\right)  p^{\alpha}(\mathbf{x})}d{\hat
{\mathbf{x}}}\right)  ^{-1}\right\rangle _{\mathbf{r}|\mathbf{x}}\right\rangle
_{\mathbf{x}}\nonumber\\
&  \leq\ln\int_{{\mathcal{R}}}p(\mathbf{r})\frac{\int_{{{\mathcal{X}}}%
}p^{\beta}\left(  \mathbf{r}|\mathbf{x}\right)  p^{\alpha}(\mathbf{x}%
)d\mathbf{x}}{\int_{{{\mathcal{X}}}}p^{\beta}(\mathbf{r}|\hat{\mathbf{x}%
})p^{\alpha}({\hat{\mathbf{x}}})d{\hat{\mathbf{x}}}}d\mathbf{r}\nonumber\\
&  =0. \label{A.Lma1aP.b}%
\end{align}

Combining Equations (\ref{A.Lma1aP.a}) and (\ref{A.Lma1aP.b}), we immediately
get the lower bound in Equation (\ref{Lma1a}).

In this section, we use integral for variable $\mathbf{x}$, although our
argument is valid for both continuous variables and discrete variables. For
discrete variables, we just need to replace each integral by a summation, and
our argument remains valid without other modification. The same is true for
the response variable $\mathbf{r}$.

To prove the upper bound, let%
\begin{equation}
\Phi\left[  q({\hat{\mathbf{x}}})\right]  =\int_{{\mathcal{R}}}p\left(
\mathbf{r}|\mathbf{x}\right)  \int_{{\mathcal{X}}}q({\hat{\mathbf{x}}}%
)\ln\left(  \frac{p\left(  \mathbf{r}|\mathbf{x}\right)  q({\hat{\mathbf{x}}%
})}{p\left(  \mathbf{r}|{\hat{\mathbf{x}}}\right)  p({\hat{\mathbf{x}}}%
)}\right)  d{\hat{\mathbf{x}}}d\mathbf{r,} \label{A.Lma1aP.1}%
\end{equation}
where $q({\hat{\mathbf{x}}})$ satisfies%
\begin{equation}
\left\{
\begin{array}
[c]{l}%
\int_{{\mathcal{X}}}q({\hat{\mathbf{x}}})d{\hat{\mathbf{x}}}=1\\
q({\hat{\mathbf{x}}})\geq0
\end{array}
\right.  . \label{A.Lma1aP.2}%
\end{equation}

By Jensen's inequality, we get%
\begin{equation}
\Phi\left[  q({\hat{\mathbf{x}}})\right]  \geq\int_{{\mathcal{R}}}p\left(
\mathbf{r}|\mathbf{x}\right)  \ln\left(  \frac{p\left(  \mathbf{r}%
|\mathbf{x}\right)  }{p\left(  \mathbf{r}\right)  }\right)  d\mathbf{r}.
\label{A.Lma1aP.3}%
\end{equation}
To find a function $q({\hat{\mathbf{x}}})$ that minimizes $\Phi\left[
q({\hat{\mathbf{x}}})\right]  $, we apply the variational principle as
follows:%
\begin{equation}
\frac{\partial\tilde{\Phi}\left[  q({\hat{\mathbf{x}}})\right]  }{\partial
q({\hat{\mathbf{x}}})}=\int_{{\mathcal{R}}}p\left(  \mathbf{r}|\mathbf{x}%
\right)  \ln\left(  \frac{p\left(  \mathbf{r}|\mathbf{x}\right)
q({\hat{\mathbf{x}}})}{p\left(  \mathbf{r}|{\hat{\mathbf{x}}}\right)
p({\hat{\mathbf{x}}})}\right)  d\mathbf{r}+1+\lambda, \label{A.Lma1aP.4}%
\end{equation}
where $\lambda$ is the Lagrange multiplier and
\begin{equation}
\tilde{\Phi}\left[  q({\hat{\mathbf{x}}})\right]  =\Phi\left[  q(\mathbf{\hat
{x}})\right]  +\lambda\left(  \int_{{\mathcal{X}}}q(\hat{\mathbf{x}}%
)d{\hat{\mathbf{x}}}-1\right)  . \label{A.Lma1aP.5}%
\end{equation}

Setting $\frac{\partial\tilde{\Phi}\left[  q({\hat{\mathbf{x}}})\right]
}{\partial q({\hat{\mathbf{x}}})}=0$ and using the constraint in Equation
(\ref{A.Lma1aP.2}), we find the optimal solution%
\begin{equation}
q^{\ast}({\hat{\mathbf{x}}})=\frac{p({\hat{\mathbf{x}}})\exp\left(  -D\left(
\mathbf{x}||{\hat{\mathbf{x}}}\right)  \right)  }{\int_{{\mathcal{X}}%
}p(\mathbf{\check{x}})\exp\left(  -D\left(  \mathbf{x|}|\mathbf{\check{x}%
}\right)  \right)  d\mathbf{\check{x}}}. \label{A.Lma1aP.6}%
\end{equation}

Thus, the variational lower bound of $\Phi\left[  q({\hat{\mathbf{x}}%
})\right]  $ is given by%
\begin{equation}
\Phi\left[  q^{\ast}({\hat{\mathbf{x}}})\right]  ={\underset{q(\mathbf{\hat
{x}})}{\min}}\,\Phi\left[  q({\hat{\mathbf{x}}})\right]  =-\ln\left(
\int_{{\mathcal{X}}}p({\hat{\mathbf{x}}})\exp\left(  -D\left(  \mathbf{x}%
||{\hat{\mathbf{x}}}\right)  \right)  d{\hat{\mathbf{x}}}\right)  d\mathbf{x}.
\label{A.Lma1aP.8}%
\end{equation}

Therefore, from Equations (\ref{MI}), (\ref{A.Lma1aP.3}) and (\ref{A.Lma1aP.8}%
), we get the upper bound in Equation (\ref{Lma1a}). This completes the proof
of {Theorem~ \ref{Theorem 1}}.%
%TCIMACRO{\TeXButton{qed}{\qed}}%
%BeginExpansion
\qed
%EndExpansion

\subsection{Proof of Theorem \ref{Theorem 2}}

\label{A.Lma1aPb}It follows from Equation (\ref{Rem1.1}) that%
\begin{align}
I_{\beta_{1},\alpha_{1}}  &  =-\left\langle \ln\left\langle \exp\left(
-\beta_{1}D_{\beta_{1}}\left(  \mathbf{x}||{\hat{\mathbf{x}}}\right)  +\left(
1-\alpha_{1}\right)  \ln\frac{p\left(  \mathbf{x}\right)  }{p\left(
{\hat{\mathbf{x}}}\right)  }\right)  \right\rangle _{{\hat{\mathbf{x}}}%
}\right\rangle _{\mathbf{x}}\nonumber\\
&  \leq-\left\langle \ln\left\langle \exp\left(  -e^{-1}D\left(
\mathbf{x}||{\hat{\mathbf{x}}}\right)  \right)  \right\rangle _{{\hat
{\mathbf{x}}}}\right\rangle _{\mathbf{x}}=I_{e}\nonumber\\
&  \leq-\left\langle \ln\left\langle \exp\left(  -D\left(  \mathbf{x}%
||{\hat{\mathbf{x}}}\right)  \right)  \right\rangle _{{\hat{\mathbf{x}}}%
}\right\rangle _{\mathbf{x}}=I_{u}, \label{Ib1<Ie}%
\end{align}
where $\beta_{1}=e^{-1}$ and $\alpha_{1}=1$. We immediately get Equation
(\ref{Ib1<Ie<Iu}). This completes the proof of {Theorem \ref{Theorem 2}}.%
%TCIMACRO{\TeXButton{qed}{\qed}}%
%BeginExpansion
\qed
%EndExpansion

\subsection{Proof of Theorem \textbf{\ref{Theorem 3}}}

\label{A.Thm1aP} For the lower bound $I_{\beta,\alpha}$, we have%
\begin{align}
I_{\beta,\alpha}  &  =-\sum\nolimits_{m=1}^{M}p(\mathbf{x}_{m})\ln\left(
\sum\nolimits_{\check{m}=1}^{M}\left(  \frac{p\left(  \mathbf{x}_{\check{m}%
}\right)  }{p\left(  \mathbf{x}_{m}\right)  }\right)  ^{\alpha}\exp\left(
-\beta D_{\beta}\left(  \mathbf{x}_{m}|\mathbf{x}_{\check{m}}\right)  \right)
\right) \nonumber\\
&  =-\sum\nolimits_{m=1}^{M}p(\mathbf{x}_{m})\ln\left(  1+d\left(
\mathbf{x}_{m}\right)  \right)  +H\left(  X\right)  , \label{A.Thm1aP.1}%
\end{align}
where%
\begin{equation}
d\left(  \mathbf{x}_{m}\right)  =\sum\nolimits_{\check{m}\in{\mathcal{M}%
}-\left\{  m\right\}  }\left(  \frac{p\left(  \mathbf{x}_{\check{m}}\right)
}{p\left(  \mathbf{x}_{m}\right)  }\right)  ^{\alpha}\exp\left(  -\beta
D_{\beta}\left(  \mathbf{x}_{m}|\mathbf{x}_{_{\check{m}}}\right)  \right)  .
\label{A.Thm1aP.1b}%
\end{equation}

Now, consider
\begin{align}
&  \ln\left(  1+d\left(  \mathbf{x}_{m}\right)  \right) \nonumber\\
&  =\ln\left(  1+a\left(  \mathbf{x}_{m}\right)  +b\left(  \mathbf{x}%
_{m}\right)  \right) \nonumber\\
&  =\ln\left(  1+a\left(  \mathbf{x}_{m}\right)  \right)  +\ln\left(
1+b\left(  \mathbf{x}_{m}\right)  \left(  1+a\left(  \mathbf{x}_{m}\right)
\right)  ^{-1}\right) \nonumber\\
&  =\ln\left(  1+a\left(  \mathbf{x}_{m}\right)  \right)  +O\left(
N^{-\gamma}\right)  , \label{A.Thm1aP.2}%
\end{align}
where
\begin{subequations}
\begin{align}
a\left(  \mathbf{x}_{m}\right)   &  =\sum\nolimits_{{\hat{m}}\in{\mathcal{M}%
}_{m}^{\beta}}\left(  \frac{p\left(  \mathbf{x}_{{\hat{m}}}\right)  }{p\left(
\mathbf{x}_{m}\right)  }\right)  ^{\alpha}\exp\left(  -\beta D_{\beta}\left(
\mathbf{x}_{m}||\mathbf{x}_{{\hat{m}}}\right)  \right)  ,\label{A.Thm1aP.2a}\\
b\left(  \mathbf{x}_{m}\right)   &  =\sum\nolimits_{\check{m}\in
{\mathcal{M-M}}_{m}^{\beta}}\left(  \frac{p\left(  \mathbf{x}_{\check{m}%
}\right)  }{p\left(  \mathbf{x}_{m}\right)  }\right)  ^{\alpha}\exp\left(
-\beta D_{\beta}\left(  \mathbf{x}_{m}||\mathbf{x}_{\check{m}}\right)  \right)
\nonumber\\
&  \leq N^{-\gamma_{1}}\sum\nolimits_{{\check{m}}\in{\mathcal{M-M}}_{m}%
^{\beta}}\left(  \frac{p\left(  \mathbf{x}_{\check{m}}\right)  }{p\left(
\mathbf{x}_{m}\right)  }\right)  ^{\alpha}=O\left(  N^{-\gamma_{1}}\right)  .
\label{A.Thm1aP.2b}%
\end{align}

Combining Equations (\ref{A.Thm1aP.1}) and (\ref{A.Thm1aP.2}) and {Theorem
\ref{Theorem 1}}, we get the lower bound in Equation (\ref{Thm1a}). In a
manner similar to the above, we can get the upper bound in
Equations~(\ref{Thm1a}) and (\ref{Thm1a.0}). This completes the proof of
{Theorem \ref{Theorem 3}}.%
%TCIMACRO{\TeXButton{qed}{\qed}}%
%BeginExpansion
\qed
%EndExpansion

\subsection{Proof of Theorem \ref{Theorem 4}}

The upper bound $I_{u}$ for mutual information $I$ in Equation (\ref{Lma1a})
can be written as
\end{subequations}
\begin{align}
I_{u}  &  =-\int_{{\mathcal{X}}}\left(  \ln\int_{{\mathcal{X}}}p\left(
{\hat{\mathbf{x}}}\right)  \exp\left(  -D\left(  \mathbf{x}|\hat{\mathbf{x}%
}\right)  \right)  d{\hat{\mathbf{x}}}\right)  p\left(  \mathbf{x}\right)
d\mathbf{x}\nonumber\\
&  =-\left\langle \ln\left(  \int_{{\mathcal{X}}}\exp\left(  \left\langle
{L(\mathbf{r}|{\hat{\mathbf{x}}})-L(\mathbf{r}|\mathbf{x})}\right\rangle
_{\mathbf{r}|\mathbf{x}}\right)  d{\hat{\mathbf{x}}}\right)  \right\rangle
_{\mathbf{x}}+H\left(  X\right)  . \label{A.Thm1AP.1}%
\end{align}
where $L\left(  \mathbf{r}|{\hat{\mathbf{x}}}\right)  =\ln\left(  p\left(
\mathbf{r}|{\hat{\mathbf{x}}}\right)  p\left(  {\hat{\mathbf{x}}}\right)
\right)  $ and $L\left(  \mathbf{r}|\mathbf{x}\right)  =\ln\left(  p\left(
\mathbf{r}|\mathbf{x}\right)  p\left(  \mathbf{x}\right)  \right)  $.

Consider the Taylor expansion for $L(\mathbf{r}|{\hat{\mathbf{x}}})$ around
$\mathbf{x}$. Assuming that $L(\mathbf{r}|{\hat{\mathbf{x}}})$ is twice
continuously differentiable for any ${\hat{\mathbf{x}}}\in{{\mathcal{X}}%
}_{\omega}(\mathbf{x})$, we get%
\begin{align}
&  \left\langle {L(\mathbf{r}|{\hat{\mathbf{x}}})-L(\mathbf{r}|\mathbf{x}%
)}\right\rangle _{\mathbf{r}|\mathbf{x}}\nonumber\\
&  ={\mathbf{y}^{T}\mathbf{v}}_{1}-{\dfrac{1}{2}\mathbf{y}^{T}\mathbf{y}%
}-{\dfrac{1}{2}\mathbf{y}^{T}}\mathbf{G}{^{-1/2}}\left(  {\mathbf{x}}\right)
\left(  \mathbf{G}(\breve{\mathbf{x}})-\mathbf{G}(\mathbf{x})\right)
\mathbf{G}{^{-1/2}}\left(  {\mathbf{x}}\right)  \mathbf{y}
\label{A.Thm1AP.1_1}%
\end{align}
where%
\begin{equation}
{\mathbf{y}}=\mathbf{G}{{^{1/2}}\left(  {\mathbf{x}}\right)  (\mathbf{\hat{x}%
}}-{\mathbf{x})}, \label{A.y}%
\end{equation}%
\begin{equation}
\mathbf{v}{_{1}}=\mathbf{G}{^{-1/2}}\left(  {\mathbf{x}}\right)  {{q}^{\prime
}(\mathbf{x})}%
\end{equation}
and%
\begin{equation}
{\breve{\mathbf{x}}}={\mathbf{x}}+t\left(  {\hat{\mathbf{x}}}-{\mathbf{x}%
}\right)  \in{{\mathcal{X}}}_{\omega}(\mathbf{x}), \quad t\in\left(
0,1\right)  . \label{A.x.t}%
\end{equation}

For later use, we also define
\begin{equation}
\mathbf{v}=\mathbf{G}{^{-1/2}}\left(  {\mathbf{x}}\right)  {l^{\prime
}(\mathbf{r}|\mathbf{x})}%
\end{equation}
where
\begin{equation}
l\left(  \mathbf{r}|\mathbf{x}\right)  =\ln p\left(  \mathbf{r}|\mathbf{x}%
\right)  .
\end{equation}

%\begin{equation}
%\left\{
%\begin{array}
%[c]{l}%
%\mathbf{\tilde{v}=\mathbf{v}}+\mathbf{\mathbf{v}}_{1}\\
%\mathbf{v}=\mathbf{G}{^{-1/2}}\left(  {\mathbf{x}}\right)  {l^{\prime
%}(\mathbf{r}|\mathbf{x})}\\
%\mathbf{v}{_{1}}=\mathbf{G}{^{-1/2}}\left(  {\mathbf{x}}\right)
%{\mathbf{q}^{\prime}(\mathbf{x})}%
%\end{array}
%\right.  \label{A.vy}%
%\end{equation}

Since $\mathbf{G}(\breve{\mathbf{x}})$ is continuous and symmetric for
$\breve{\mathbf{x}}\in{{\mathcal{X}}}$, for any $\epsilon\in\left(
0\text{,\thinspace}1\right)  $, there is a ${\varepsilon}\in\left(
0,\text{\thinspace}\omega\right)  $\ such that%
\begin{equation}
\left\vert {\mathbf{y}^{T}\mathbf{G}{^{-1/2}}\left(  {\mathbf{x}}\right)
\left(  \mathbf{G}(\breve{\mathbf{x}})-\mathbf{G}(\mathbf{x})\right)
\mathbf{G}{^{-1/2}}\left(  {\mathbf{x}}\right)  \mathbf{y}}\right\vert
<\epsilon{\left\Vert {\mathbf{y}}\right\Vert ^{2}} \label{A.Thm1AP.1_2a}%
\end{equation}
for all ${\mathbf{y}}\in{\mathcal{Y}_{\varepsilon}}$, where ${\mathcal{Y}%
_{\varepsilon}}=\left\{  {\mathbf{y}}\in%
%TCIMACRO{\U{211d} }%
%BeginExpansion
\mathbb{R}
%EndExpansion
^{K}:\left\Vert \mathbf{y}\right\Vert <\varepsilon\sqrt{N}\right\}  $. Then,
we get%
\begin{align}
&  \ln\left(  \int_{{\mathcal{X}}}\exp\left(  \left\langle {L(\mathbf{r}%
|{\hat{\mathbf{x}}})-L(\mathbf{r}|\mathbf{x})}\right\rangle _{\mathbf{r}%
|\mathbf{x}}\right)  d{\hat{\mathbf{x}}}\right) \nonumber\\
&  \geq-\dfrac{1}{2}\ln\left(  \det\left(  \mathbf{G}(\mathbf{x})\right)
\right)  +\ln\int_{{\mathcal{Y}_{\varepsilon}}}\exp\left(  {\mathbf{y}%
^{T}\mathbf{v}}_{1}-{\dfrac{1}{2}}\left(  1+\epsilon\right)  {\mathbf{y}%
^{T}\mathbf{y}}\right)  d{\mathbf{y}} \label{A.Thm1AP.1_3}%
\end{align}
and with Jensen's inequality,%
\begin{align}
&  \ln\int_{{\mathcal{Y}_{\varepsilon}}}\exp\left(  {\mathbf{y}^{T}\mathbf{v}%
}_{1}-{\dfrac{1}{2}}\left(  1+\epsilon\right)  {\mathbf{y}^{T}\mathbf{y}%
}\right)  d{\mathbf{y}}\nonumber\\
&  \geq\ln{\Psi}_{\varepsilon}+\int_{{\mathcal{\hat{Y}}_{\varepsilon}}%
}{\mathbf{y}^{T}\mathbf{v}}_{1}\phi_{\varepsilon}\left(  \mathbf{y}\right)
d{\mathbf{y}}\nonumber\\
&  =\frac{K}{2}\ln\left(  \frac{2\pi}{1+\epsilon}\right)  +O\left(
N^{-K/2}e^{-N\delta}\right)  , \label{A.Thm1AP.1_4}%
\end{align}
where $\delta$ is a positive constant, $\int_{{\mathcal{\hat{Y}}_{\varepsilon
}}}{\mathbf{y}^{T}\mathbf{v}}_{1}\phi_{\varepsilon}\left(  \mathbf{y}\right)
d{\mathbf{y=0}}$,
\begin{equation}
\left\{
\begin{array}
[c]{l}%
\phi_{\varepsilon}\left(  \mathbf{y}\right)  ={\Psi}_{\varepsilon}^{-1}%
{\exp\left(  -\dfrac{1}{2}\left(  1+\epsilon\right)  \mathbf{y}^{T}%
\mathbf{y}\right)  }\\
\Psi_{\varepsilon}={\int_{{\mathcal{\hat{Y}}_{\varepsilon}}}\exp\left(
-\dfrac{1}{2}\left(  1+\epsilon\right)  \mathbf{y}^{T}\mathbf{y}\right)
d}\mathbf{y}%
\end{array}
\right.  \label{A.Thm1AP.1_4a}%
\end{equation}
and%
\begin{equation}
{\mathcal{\hat{Y}}_{\varepsilon}}={\left\{  \mathbf{y}\in%
%TCIMACRO{\U{211d} }%
%BeginExpansion
\mathbb{R}
%EndExpansion
^{K}:\left\vert y_{k}\right\vert <\varepsilon\sqrt{N/K},\text{\ }%
k=1,2,\cdots,K\right\}  }\subseteq{\mathcal{Y}_{\varepsilon}.}
\label{A.Thm1AP.1_4b}%
\end{equation}

Now, we evaluate
\begin{align}
\Psi_{\varepsilon}  &  ={\int_{%
%TCIMACRO{\U{211d} }%
%BeginExpansion
\mathbb{R}
%EndExpansion
^{K}}\exp\left(  -\dfrac{1}{2}\left(  1+\epsilon\right)  \mathbf{y}%
^{T}\mathbf{y}\right)  d}\mathbf{y}-{\int_{%
%TCIMACRO{\U{211d} }%
%BeginExpansion
\mathbb{R}
%EndExpansion
^{K}-{\mathcal{\hat{Y}}_{\varepsilon}}}\exp\left(  -\dfrac{1}{2}\left(
1+\epsilon\right)  \mathbf{y}^{T}\mathbf{y}\right)  d}\mathbf{y}\nonumber\\
&  =\left(  \frac{2\pi}{1+\epsilon}\right)  ^{K/2}-{\int_{%
%TCIMACRO{\U{211d} }%
%BeginExpansion
\mathbb{R}
%EndExpansion
^{K}-{\mathcal{\hat{Y}}_{\varepsilon}}}\exp\left(  -\dfrac{1}{2}\left(
1+\epsilon\right)  \mathbf{y}^{T}\mathbf{y}\right)  d}\mathbf{y}.
\label{A.Thm1AP.1_5}%
\end{align}

Performing integration by parts with $\int_{a}^{\infty}e^{-t^{2}/2}%
dt=\frac{e^{-a^{2}/2}}{a}-\int_{a}^{\infty}\frac{e^{-t^{2}/2}}{t^{2}}dt$, we
find
\begin{align}
{\int_{%
%TCIMACRO{\U{211d} }%
%BeginExpansion
\mathbb{R}
%EndExpansion
^{K}-{\mathcal{\hat{Y}}_{\varepsilon}}}\exp\left(  -\dfrac{1}{2}\left(
1+\epsilon\right)  \mathbf{y}^{T}\mathbf{y}\right)  d}\mathbf{y}  &  \leq
\frac{{\exp\left(  -\dfrac{1}{2}\left(  1+\epsilon\right)  \varepsilon
^{2}N\right)  }}{\left(  \left(  1+\epsilon\right)  ^{2}\varepsilon
^{2}N/\left(  4K\right)  \right)  ^{K/2}}\nonumber\\
&  =O\left(  N^{-K/2}e^{-N\delta}\right)  , \label{A.Thm1AP.1_6}%
\end{align}
for some constant $\delta>0$.

Combining Equations (\ref{A.Thm1AP.1}), (\ref{A.Thm1AP.1_3}) and
(\ref{A.Thm1AP.1_4}), we get%
\begin{equation}
I_{u}\leq\dfrac{1}{2}\left\langle \ln\left(  \det\left(  \frac{\left(
1+\epsilon\right)  }{2\pi}\mathbf{G}(\mathbf{x})\right)  \right)
\right\rangle _{\mathbf{x}}+H\left(  X\right)  +O\left(  N^{-K/2}e^{-N\delta
}\right)  . \label{A.Thm1AP.1_7}%
\end{equation}

On the other hand, from Equation (\ref{A.Thm1AP.1_2a}) and the condition in
Equation (\ref{Thm1A.1a}), we obtain
\begin{align}
&  \int_{{\mathcal{X}}_{\varepsilon}\left(  {\mathbf{x}}\right)  }\exp\left(
\left\langle {L(\mathbf{r}|{\hat{\mathbf{x}}})-L(\mathbf{r}|\mathbf{x}%
)}\right\rangle _{\mathbf{r}|\mathbf{x}}\right)  d{\hat{\mathbf{x}}%
}\nonumber\\
&  \leq\det\left(  \mathbf{G}(\mathbf{x})\right)  ^{-1/2}\int_{%
%TCIMACRO{\U{211d} }%
%BeginExpansion
\mathbb{R}
%EndExpansion
^{K}}\exp\left(  {\mathbf{y}^{T}\mathbf{v}}_{1}-{\dfrac{1}{2}}\left(
1-\epsilon\right)  {\mathbf{y}^{T}\mathbf{y}}\right)  d{\mathbf{y}}\nonumber\\
&  =\det\left(  \frac{1-\epsilon}{2\pi}\mathbf{G}(\mathbf{x})\right)
^{-1/2}\exp\left(  {\dfrac{1}{2}}\left(  1-\epsilon\right)  ^{-1}{\mathbf{v}%
}^{T}{\mathbf{v}}_{1}\right)  \label{A.Thm1AP.2_3}%
\end{align}
and%
\begin{align}
&  \int_{{\mathcal{X}}}\exp\left(  \left\langle {L(\mathbf{r}|\hat{\mathbf{x}%
})-L(\mathbf{r}|\mathbf{x})}\right\rangle _{\mathbf{r}|\mathbf{x}}\right)
d{\hat{\mathbf{x}}}\nonumber\\
&  =\int_{{\mathcal{X}}_{\varepsilon}\left(  {\mathbf{x}}\right)  }\exp\left(
\left\langle {L(\mathbf{r}|{\hat{\mathbf{x}}})-L(\mathbf{r}|\mathbf{x}%
)}\right\rangle _{\mathbf{r}|\mathbf{x}}\right)  d{\hat{\mathbf{x}}}%
+\int_{{\mathcal{X-}{\mathcal{X}}}_{\varepsilon}\left(  {\mathbf{x}}\right)
}\exp\left(  \left\langle {L(\mathbf{r}|{\hat{\mathbf{x}}})-L(\mathbf{r}%
|\mathbf{x})}\right\rangle _{\mathbf{r}|\mathbf{x}}\right)  d\hat{\mathbf{x}%
}\nonumber\\
&  \leq\det\left(  \frac{1-\epsilon}{2\pi}\mathbf{G}(\mathbf{x})\right)
^{-1/2}\left(  \exp\left(  {\dfrac{{\mathbf{v}}^{T}{\mathbf{v}}_{1}}{2\left(
1-\epsilon\right)  }}\right)  +O\left(  N^{-1}\right)  \right)  .
\label{A.Thm1AP.2_4}%
\end{align}

It follows from Equations (\ref{A.Thm1AP.1}) and (\ref{A.Thm1AP.2_4}) that
\begin{align}
&  \left\langle \ln\left(  \int_{{\mathcal{X}}}\exp\left(  \left\langle
{L(\mathbf{r}|{\hat{\mathbf{x}}})-L(\mathbf{r}|\mathbf{x})}\right\rangle
_{\mathbf{r}|\mathbf{x}}\right)  d{\hat{\mathbf{x}}}\right)  \right\rangle
_{\mathbf{x}}\nonumber\\
&  \leq-\dfrac{1}{2}\left\langle \ln\left(  \det\left(  \frac{\left(
1-\epsilon\right)  }{2\pi}\mathbf{G}(\mathbf{x})\right)  \right)
\right\rangle _{\mathbf{x}}+{\dfrac{1}{2}}\left(  1-\epsilon\right)
^{-1}\left\langle {\mathbf{v}}^{T}{\mathbf{v}}_{1}\right\rangle _{\mathbf{x}%
}+O\left(  N^{-1}\right)  . \label{A.Thm1AP.2_5}%
\end{align}

Note that$\ $%
\begin{equation}
\left\langle {\mathbf{v}}^{T}{\mathbf{v}}_{1}\right\rangle _{\mathbf{x}%
}=O\left(  N^{-1}\right)  . \label{A.Thm1AP.2_6}%
\end{equation}

Now, we have%
\begin{equation}
I_{u}\geq\dfrac{1}{2}\left\langle \ln\left(  \det\left(  \frac{\left(
1-\epsilon\right)  }{2\pi}\mathbf{G}(\mathbf{x})\right)  \right)
\right\rangle _{\mathbf{x}}+H\left(  X\right)  +O\left(  N^{-1}\right)  .
\label{A.Thm1AP.2_7}%
\end{equation}

Since $\epsilon$ is arbitrary, we can let it go to zero. Therefore, from
Equations (\ref{Lma1a}), (\ref{A.Thm1AP.1_7}) and (\ref{A.Thm1AP.2_7}), we
obtain the upper bound in Equation (\ref{Thm1A}).

The Taylor expansion of $h\left(  {\mathbf{\hat{x}}},{\mathbf{x}}\right)
=\left\langle \left(  \frac{p(\mathbf{r}|{\hat{\mathbf{x}}})}{p\left(
\mathbf{r}|\mathbf{x}\right)  }\right)  ^{\beta}\right\rangle _{\mathbf{r}%
|\mathbf{x}}$ around $\mathbf{x}$ is
\begin{align}
h\left(  {\mathbf{\hat{x}}},{\mathbf{x}}\right)   &  =1+\left\langle
\frac{\beta}{p(\mathbf{r}|\mathbf{x})}\frac{\partial p(\mathbf{r}|\mathbf{x}%
)}{\partial\mathbf{x}}\right\rangle _{\mathbf{r}|\mathbf{x}}{(\mathbf{\hat{x}%
}}-{\mathbf{x})+}\nonumber\\
&  {(\mathbf{\hat{x}}}-{\mathbf{x})}^{T}\left\langle \frac{\beta
}{2p(\mathbf{r}|\mathbf{x})^{2}}\left(  \left(  \beta-1\right)  \frac{\partial
p(\mathbf{r}|\mathbf{x})}{\partial\mathbf{x}}\frac{\partial p(\mathbf{r}%
|\mathbf{x})}{\partial\mathbf{x}^{T}}+p(\mathbf{r}|\mathbf{x})\frac
{\partial^{2}p(\mathbf{r}|\mathbf{x})}{\partial\mathbf{x}\partial
\mathbf{x}^{T}}\right)  \right\rangle _{\mathbf{r}|\mathbf{x}}{(\mathbf{\hat
{x}}}-{\mathbf{x})+\cdots}\nonumber\\
&  =1-\frac{\beta\left(  1-\beta\right)  }{2}{(\mathbf{\hat{x}}}-{\mathbf{x}%
)}^{T}\mathbf{J}(\mathbf{x}){(\mathbf{\hat{x}}}-{\mathbf{x})+\cdots.}
\label{hxx}%
\end{align}

In a similar manner as described above, we obtain the asymptotic relationship
(\ref{Ibeta1}):
\begin{align}
I_{\beta,\alpha}  &  =I_{\gamma}+O\left(  N^{-1}\right) \nonumber\\
&  =\frac{1}{2}\int_{{{\mathcal{X}}}}p(\mathbf{x})\ln\left(  \det\left(
\frac{\mathbf{G}_{\gamma}(\mathbf{x})}{2\pi}\right)  \right)  d\mathbf{x}%
+H(X). \label{A.Thm1AP.3}%
\end{align}

Notice that $0<\gamma=\beta\left(  1-\beta\right)  \leq1/4$ and the equality
holds when $\beta=\beta_{0}=1/2$. Thus, we have
\begin{equation}
\det\left(  \mathbf{G}_{\gamma}(\mathbf{x})\right)  \leq\det\left(
\mathbf{G}_{\gamma_{0}}(\mathbf{x})\right)  . \label{A.Thm1AP.3_0a}%
\end{equation}

Combining Equations (\ref{Lma1a}), (\ref{A.Thm1AP.3}) and (\ref{A.Thm1AP.3_0a}%
) yields the lower bound in Equation (\ref{Thm1A}).

The proof of Equation (\ref{Ie=Ia}) is similar. This completes the proof of
{Theorem \ref{Theorem 4}}.%
%TCIMACRO{\TeXButton{qed}{\qed}}%
%BeginExpansion
\qed
%EndExpansion

%\phantomsection\addcontentsline{toc}{section}{References}
%\reftitle{References} \externalbibliography{yes}
\phantomsection\addcontentsline{toc}{section}{References}
\bibliographystyle{apalike2}
\bibliography{EN091718}

\end{document}